\documentclass{aa}  

\usepackage{graphicx}
\usepackage{txfonts}
\usepackage{lipsum}
\usepackage{color}
\usepackage[dvipsnames]{xcolor}
\usepackage{ulem}
\usepackage{siunitx}
\usepackage{ragged2e} 
\usepackage{multirow}
\usepackage{siunitx}
\usepackage{booktabs}
\usepackage{natbib}
\usepackage{bm}
\usepackage[colorlinks=true,citecolor=blue,linkcolor=red,filecolor=magenta,urlcolor=cyan]{hyperref}
\usepackage{subcaption}         
\usepackage{lscape}             
\usepackage{placeins}          
\makeatletter

\newcommand{\Rmnum}[1]{\expandafter\@slowromancap\romannumeral #1@}
\makeatother

\def\neworcid#1{\kern .08em\href{https://orcid.org/#1}{\includegraphics[keepaspectratio,width=0.7em]{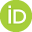}}}

\begin{document}

   \title{Optical transmission spectrum of HAT-P-47b:\\ evidence for aerosols and tentative TiO absorption}

   \subtitle{}

   \titlerunning{Transmission spectra of HAT-P-47b}

   \author{Wan-Hao Wang\inst{\ref{pmo},\ref{ustc}}\neworcid{0009-0008-1371-229X} 
   \and Guo Chen\inst{\ref{pmo}}\corrauth{guochen@pmo.ac.cn}\neworcid{0000-0003-0740-5433} 
   \and Fei Yan\inst{\ref{ustc-astro}}\neworcid{0000-0001-9585-9034}
   \and Chengzi Jiang\inst{\ref{iac},\ref{ull}}\neworcid{0000-0003-1381-5527}
   \and Luigi Mancini\inst{\ref{uroma},\ref{oato}}\neworcid{0000-0002-9428-8732}
   \and Enric Pall\'e\inst{\ref{iac},\ref{ull}}\neworcid{0000-0003-0987-1593}
   \and Felipe Murgas\inst{\ref{iac},\ref{ull}}\neworcid{0000-0001-9087-1245}
   \and Hannu Parviainen\inst{\ref{ull},\ref{iac}}\neworcid{0000-0001-5519-1391}
          }

   \institute{
   CAS Key Laboratory of Planetary Sciences, Purple Mountain Observatory, Chinese Academy of Sciences, Nanjing, 210023, PR China\label{pmo}
   \and School of Astronomy and Space Science, University of Science and Technology of China, Hefei, 230026, PR China\label{ustc}
   \and Department of Astronomy, University of Science and Technology of China, Hefei 230026, PR China\label{ustc-astro}
   \and Instituto de Astrof\'isica de Canarias (IAC), V\'ia L\'actea s/n, 38205 La Laguna, Tenerife, Spain\label{iac}
   \and Departamento de Astrof\'isica, Universidad de La Laguna (ULL), C/ Padre Herrera, 38206 La Laguna, Tenerife, Spain\label{ull}
   \and Department of Physics, University of Rome ``Tor Vergata'', Via della Ricerca Scientifica 1, 00133 Rome, Italy\label{uroma}
   \and INAF -- Turin Astrophysical Observatory, via Osservatorio 20, 10025 Pino Torinese, Italy\label{oato}
   }

   \date{Received April 3, 2026}
 
  \abstract
  % context heading (optional)
  % {} leave it empty if necessary  
   {Transmission spectroscopy enables the characterization of exoplanet atmospheres by probing absorption features in their terminator regions. In the optical, it is particularly sensitive to metal oxides and atomic species that can strongly influence atmospheric energy balance and thermal structure. }
  % aims heading (mandatory)
   {We aim to investigate the atmospheric properties of the hot Jupiter HAT-P-47b through optical transmission spectroscopy. }
  % methods heading (mandatory)
   {Thirteen TESS transits were analyzed to refine the planetary ephemeris and system parameters. Two ground-based transits were observed with the Multi-Object Double Spectrograph (MODS) on the Large Binocular Telescope (LBT) and the Optical System for Imaging and Low-Intermediate-Resolution Integrated Spectroscopy+ (OSIRIS+) on the Gran Telescopio Canarias (GTC). Chromatic transit light curves were modeled to derive instrument-specific transmission spectra and multiple Bayesian spectral retrievals were performed to characterize the atmospheric properties. }
  % results heading (mandatory)
   {The MODS transmission spectrum provides moderate Bayesian evidence ($\Delta\ln\mathcal{Z}=2.68$) for TiO absorption, whereas the OSIRIS+ spectrum does not yield statistically significant evidence for any specific opacity source. Both datasets exhibit a wavelength-dependent slope indicative of enhanced aerosol scattering. The MODS and OSIRIS+ joint free-chemistry retrieval, dominated by the higher signal-to-noise MODS data, yields moderate evidence ($\Delta\ln\mathcal{Z}=3.44$) for TiO with a log mass fraction of $-6.86^{+0.64}_{-0.63}$~dex. The same model indicates an aerosol contribution to the optical scattering opacity approximately $5000\times$ larger than pure H$_2$ Rayleigh scattering.} 
  % conclusions heading (optional), leave it empty if necessary
   {HAT-P-47b appears to host a cloudy atmosphere with evidence for aerosols and tentative evidence for TiO absorption. Future high-precision observations will be essential to confirm the presence of TiO and further characterize its atmospheric structure. }


   \maketitle

\section{Introduction}
\label{Section:introduction}
Transiting exoplanets partially block the stellar disk along the observer's line of sight, causing the transmitted starlight to be modulated by the wavelength-dependent opacities of atmospheric constituents. The resulting transit depth variations, referred to as transmission spectroscopy \citep{Seager2000}, provide insights into the chemical composition and thermal structure of the planet's day-night terminator region and impose important constraints on planetary formation and evolutionary histories \citep{Madhusudhan2012,Mordasini2016,Chachan2023,Schleich2024}. Certain metals, metal oxides, and hydrides can act as optical absorbers \citep{Sharp2007}, contributing significant opacity in the optical and near-infrared wavelengths and strongly depositing stellar energy at high altitudes \citep{Hubeny2003,Guillot2010}. 

Among these optical absorbers, titanium- and vanadium-bearing oxides have long been of particular interest. TiO and VO were predicted to remain in the gas phase in sufficiently hot, highly irradiated atmospheres and to absorb stellar radiation at low pressures, potentially producing thermal inversions in hot and ultra-hot Jupiters \citep{Fortney2008,Spiegel2009}. However, observational evidence for TiO in transmission or emission spectroscopy has been reported only for a limited number of planets, e.g., WASP-33b \citep[2710~K;][]{Nugroho2017,Cont2021}, WASP-189b \citep[2641~K;][]{Prinoth2022}, WASP-121b \citep[2409~K;][]{Ouyang2023b}, WASP-19b \citep[2077~K;][]{Sedaghati2017,Sedaghati2021}, HAT-P-41b \citep[1937~K;][]{Jiang2024}, HAT-P-65b \citep[1818~K;][]{Chen2021}. Several systems also show conflicting results between different observations or retrieval approaches. For example, TiO and VO have not been detected in high-resolution spectroscopy of WASP-121b \citep{Merritt2020}, while other studies favor alternative explanations such as hazes, SH absorption, or atomic metals for the optical opacity \citep{Evans2018,Wilson2021}. Similarly, the atmosphere of WASP-19b has been interpreted both as showing TiO absorption \citep{Sedaghati2017} and as exhibiting weak or featureless optical spectra inconsistent with strong TiO opacity \citep{Huitson2013,Espinoza2019}. Consistently, population-level analyses based on optical geometric albedos suggest that atmospheres with strong TiO absorption are uncommon \citep{Jones2026}. 

The absence of TiO, or its presence in low abundance, even in atmospheres that are highly irradiated and subject to thermal inversions, suggests that processes such as condensation, cold trapping, rainout, thermal dissociation, or other disequilibrium processes may efficiently deplete titanium-bearing species from observable regions of the atmosphere \citep{Hubeny2003,Spiegel2009,Knutson2010,Lothringer2018,Parmentier2013,Parmentier2018,Pelletier2026a}. Recent observational studies of ultra-hot Jupiters have reported evidence for depletion of refractory species in highly irradiated atmospheres relative to equilibrium expectations, particularly in the planetary terminator regions probed by transmission spectroscopy \citep{Merritt2020,Hoeijmakers2020,Maguire2023,Gandhi2023,Pelletier2023,Hoeijmakers2024,Prinoth2025}.

Clouds and aerosols provide another key component in the interpretation of optical transmission spectra. High-altitude aerosol particles can produce Rayleigh- or Mie-like scattering slopes and mute molecular absorption features. This introduces degeneracies between atmospheric composition, cloud properties, and reference pressure \citep{Wakeford2015,Pinhas2017,Gao2021}. At hot-Jupiter temperatures, aerosols may arise from condensate clouds composed of silicates, oxides, or sulfide minerals, while photochemical hazes may also form at lower pressures under strong irradiation and efficient vertical mixing \citep{Kawashima2019,Ohno2020,Gao2021}. These processes are particularly relevant for low-gravity, highly irradiated planets, where large atmospheric scale heights make both scattering slopes and weak molecular features accessible to transmission spectroscopy.

HAT-P-47b \citep{Bakos2016} is a particularly interesting target for probing the interplay between aerosol scattering and refractory optical absorbers near the Ti-bearing condensation boundary. It is a highly inflated hot sub-Saturn that orbits an F-type star with a period of $4.732182\pm 0.000013$~days, placing it on the edge of the short-period Neptunian desert \citep{Szabo2011,Mazeh2016}. The host star has a mass of $1.387\pm 0.038$~$M_\odot$, a radius of $1.515\pm 0.040$~$R_\odot$, an effective temperature of $6703\pm 50$~K, and an age of $1.5\pm 0.3$~Gyr. The planet has a high equilibrium temperature of $1605\pm 22$~K and a low bulk density, with a mass of $0.206\pm 0.039$~$M_{\rm Jup}$ and a radius of $1.313\pm 0.045$~$R_{\rm Jup}$. These properties result in a large atmospheric scale height ($H/R_\star = (1.87 \pm 0.38) \times 10^{-3}$), making it a compelling target for transmission spectroscopy. 

In this work, we present ground-based transit observations of HAT-P-47b obtained with the Multi-Object Double Spectrograph (MODS; \citealt{Pogge2010}) on the Large Binocular Telescope (LBT) and the Optical System for Imaging and Low-Intermediate-Resolution Integrated Spectroscopy+ (OSIRIS+; \citealt{Cepa2000}) on the Gran Telescopio Canarias (GTC). In addition, we incorporate space-based photometric observations acquired with the Transiting Exoplanet Survey Satellite (TESS; \citealt{Ricker2015}). Using these data, we construct the first optical transmission spectrum of HAT-P-47b to identify absorption and scattering signatures, investigate the presence of potential optical absorbers, and constrain the planet’s atmospheric properties through atmospheric retrieval analyses. 

The paper is organized as follows. Section~\ref{Section:observation} summarizes the observations and data reduction. Section~\ref{Section:lc} details the transit light-curve analysis. The transmission spectra and atmospheric retrieval framework are presented in Section~\ref{Section:retrieval}. Section~\ref{Section:discussion} discusses the atmospheric properties of HAT-P-47b and possible alternative explanations. Section~\ref{Section:conclusions} summarizes the main results.

\section{Observations and data reduction}
\label{Section:observation}

\begin{table}%[htb]
\small
\centering
\renewcommand\arraystretch{1.5} 
\caption{Observation summary.} 
\label{tab:obs_sum}
\begin{tabular}{l c}
\hline\hline\noalign{\smallskip}
\multicolumn{2}{c}{Transit \#1} \\
\hline
Instrument & LBT/MODS \\
Program ID & LBT-2019A-I0062-0 \\
PI & Luigi Mancini \\
Date (UT) & 2018-11-12 \\
UT range & 02:47 -- 07:35 \\
Airmass \tablefootmark{a} & 1.46 -- 1.00 -- 1.03\\
Seeing (arcsec) \tablefootmark{b} & 0.86 -- 4.14 (1B) \\
 & 0.76-- 3.05 (1R) \\
 & 1.08 -- 3.18 (2B) \\
Exposure time (s) & 60 (1B), 30 (1R), 45 (2B)\\
Useful exposures & 132 (1B), 147 (1R), 145 (2B)  \\
\hline\noalign{\smallskip}
\multicolumn{2}{c}{Transit \#2} \\
\hline
Instrument & GTC/OSIRIS+ \\
Program ID & GTCMULTIPLE2I-23B \\
PI & Felipe Murgas \\
Date (UT) & 2023-11-26 \\
UT range & 21:00 -- 03:00$^{+1}$ \\
Airmass \tablefootmark{a} & 1.17 -- 1.01 -- 1.34 \\
Seeing (arcsec) \tablefootmark{b} & 1.19 -- 3.18 \\
Exposure time (s) & 8.5 \\
Useful exposures & 488 \\
\noalign{\smallskip}
\hline
\end{tabular}
\tablefoot{
\tablefoottext{a}{Start – minimum – end.}
\tablefoottext{b}{95\% interval of the seeing distribution measured from the FWHM of stellar PSFs at the central wavelengths of 486~nm (1B), 746~nm (1R), 521~nm (2B), and 572~nm (GTC).}
}
\end{table}

\begin{figure}%[htb]
\centering
\includegraphics[width=1.0\hsize]{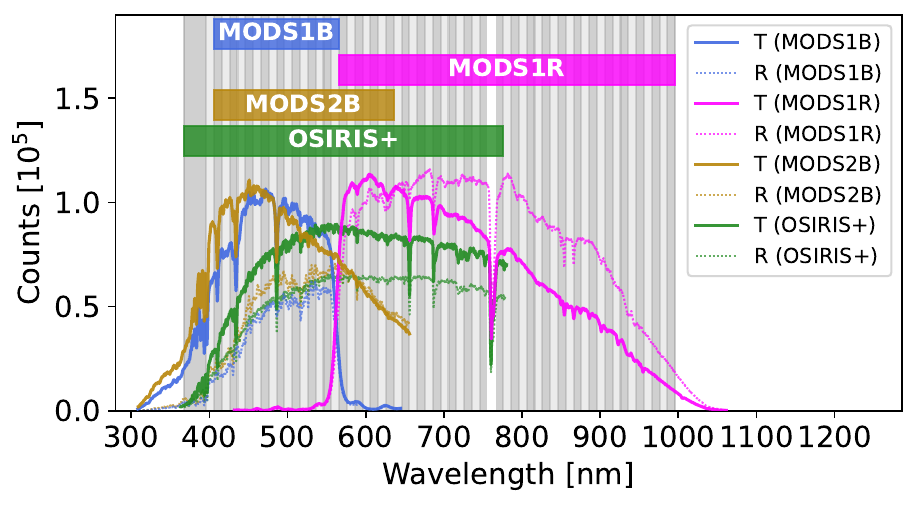}
\caption{Example stellar spectra of HAT-P-47 (T; solid line) and its reference star (R; dotted line), observed with LBT/MODS1B (blue), LBT/MODS1R (magenta), LBT/MODS2B (yellow), and GTC/OSIRIS+ (green). The vertical lines with gray-shaded regions indicate the adopted spectroscopic passbands. }
\label{fig:t&r_flux}
\end{figure}

\subsection{LBT/MODS transit spectroscopy}
We observed one transit of HAT-P-47b on 12 November 2018 with LBT/MODS. MODS1 and MODS2 are a pair of low- to medium-resolution optical spectrographs and imagers mounted on the twin 8.4-m diameter mirrors of the LBT. Each instrument covers a wavelength range of 320--1000~nm and provides a field of view (FOV) of $6^{\prime}\times 6^{\prime}$. The gratings deliver a resolving power of $\sim$2000 when operated with a narrow longslit. Both channels are equipped with e2v $\rm 3k\times 8k$ CCD detectors and $2\times 2$ pixel binning was applied during the observations. 

A custom multi-object-spectroscopy (MOS) mask was designed to enable simultaneous observations of the target star (HAT-P-47, 2MASS 02331396+3021377) and a reference star (2MASS 02332745+3022592). The mask consisted of two wide slits, each with a width of $15^{\prime\prime}$ and a length of $60^{\prime\prime}$, one centered on the target and the other on the reference star. The target and reference are separated by $3.11^{\prime}$, with $r$-band magnitudes of 10.514 and 10.475, respectively \citep{Zacharias2012}. 
MODS1 was operated in dual-channel mode, in which a dichroic splits the incoming beam at $\sim$565~nm into blue- (MODS1B) and red-optimized (MODS1R) spectrograph channels. MODS2 was used in blue-only mode (MODS2B), as the red channel (MODS2R) was unavailable during the observation. We employed the G400L grating for both MODS1B and MODS2B, and the G470L grating for MODS1R. 

The observations were conducted from UT 02:47 to 07:35 with exposure times of 60~s, 30~s and 45~s for MODS1B, MODS1R and MODS2B, respectively. No spectra were recorded for MODS1R between UT 03:16--04:06 or for MODS2B between UT 03:39--04:08. MODS1R exposures taken between UT 02:55--03:16 were discarded due to readout errors. In addition, 12, 18 and 15 exposures from MODS1B, MODS1R and MODS2B, respectively, obtained between UT 06:14--06:37, were discarded due to a systematic wavelength drift likely caused by a displacement of the stellar trace across the slit. Table~\ref{tab:obs_sum} summarizes the observational conditions. The raw data are publicly available through the LBT Archive\footnote{\url{https://archive.lbto.org/}}.

The MODS data were reduced following the procedures described in our previous work \citep{Wang2026}. Bias and flat calibrations were carried out using the \texttt{IRAF} package \citep{Tody1993} in combination with the \texttt{modsCCDRed} pipeline \citep{richard_pogge_2019_2647501}. Wavelength calibration was performed using Ne, HgAr and XeKr arc lamps with a $1.2^{\prime\prime}$-wide slit. Optimal extraction aperture radii of 16.5, 16.0, and 20.0~pixels were adopted for MODS1B, MODS1R, and MODS2B, respectively, by minimizing the root-mean-square (RMS) scatter of the white light curves. Customized \texttt{IDL} routines were developed to generate spectral cubes for the target and reference stars. 

The white light curves were constructed over 406--566~nm (MODS1B), 566--926~nm (MODS1R), and 406--566~nm (MODS2B). The telluric oxygen A band (756--766~nm) was excluded due to its low signal-to-noise ratio. Spectroscopic light curves were generated in 10~nm bins, yielding 16, 43, and 23 passbands over the wavelength ranges 406--566~nm (MODS1B), 566--996~nm (MODS1R), and 406--636~nm (MODS2B), respectively. Figure~\ref{fig:t&r_flux} shows example 1D stellar spectra of the target and reference stars, together with the adopted passbands. 

\subsection{GTC/OSIRIS+ transit spectroscopy}
We observed one transit of HAT-P-47b on 26 November 2023 with GTC/OSIRIS+ using the R1000B grism. The OSIRIS+ instrument mounted at the Cassegrain focal station is a comprehensive upgrade of the original OSIRIS that became operational on 1 January 2023. It provides an unvignetted FOV of $7.8^{\prime}\times 7.8^{\prime}$ and a pixel scale of $0.254^{\prime\prime}$ with $2\times 2$ pixel binning. The upgraded detector consists of a $\rm 4k\times 4k$ deep-depleted E2V CCD231-84, characterized by reduced fringing and improved sensitivity at the blue end of the optical wavelength range. 

Simultaneous spectroscopic observations of HAT-P-47 and a reference star (2MASS 02333866+3024167) were obtained using a $12^{\prime\prime}$-wide and $7.1^{\prime}$-long slit. The reference star has an $r$-band magnitude of 10.776 and lies $5.95^{\prime}$ from the target \citep{Zacharias2012}. The observations were conducted from UT 21:00 to 03:00$^{+1}$ with an exposure time of 8.5~s. A total of 89 exposures (UT 22:59--23:58) and 43 exposures (after UT 02:35$^{+1}$) were discarded owing to significant flux loss caused by clouds. A summary of the OSIRIS+ observing conditions is listed in Table~\ref{tab:obs_sum}. The raw data are publicly available through the GTC Archive\footnote{\url{https://gtc.sdc.cab.inta-csic.es/gtc/index.jsp}}.

The OSIRIS+ data were reduced following \citet{Wang2026}. Spectra were extracted with an optimal aperture radius of 13.5~pixels. The white light curve was generated over 368--776~nm, excluding the telluric oxygen A band (756--766~nm). Spectroscopic light curves were derived using 38 bins, primarily 10~nm wide, with two broader bins (28 and 20~nm) at the blue end to account for lower flux. Figure~\ref{fig:t&r_flux} shows example OSIRIS+ spectra and adopted bins compared to those of MODS. 

\subsection{TESS photometry}
HAT-P-47 (TIC 73448352) was observed by TESS in a 0.6--1.0~$\rm\mu m$ band during Sectors 18, 58, and 85 (S18, S58, and S85), covering 13 transits in total. In S18 (3--27 November 2019), five transits were recorded at 1800~s cadence, while four transits each were observed in S58 (29 October -- 26 November 2022) and S85 (26 October -- 21 November 2024) at 120~s cadence. 

Full frame images (FFIs) from all three sectors were processed with the Quick Look Pipeline (QLP; \citealt{Huang2020}). FFIs from S18 and S58 were also reduced using the TESS-SPOC pipeline \citep{Caldwell2020}, while target pixel files (TPFs) for S58 and S85 were produced by the SPOC pipeline \citep{Jenkins2016}. Both pipelines are operated at the Science Processing Operations Center (NASA Ames Research Center) and share a common codebase, providing simple aperture photometry (SAP; \citealt{Twicken2010,Mousis2020}) and Presearch Data Conditioned SAP (PDCSAP; \citealt{Smith2012,Stumpe2012,Stumpe2014}) light curves. 

We retrieved the TESS-SPOC light curves for S18 and the SPOC light curves for S58 and S85 from the Mikulski Archive for Space Telescopes (MAST) using the \texttt{lightkurve} Python package \citep{LightkurveCollaboration2018}. The PDCSAP fluxes were adopted for transit fitting (Section~\ref{subsect:TESS-lc}), as they were corrected for systematics and long-term trends. The dilution effects of all nearby contaminating sources have already been accounted for in the PDCSAP light curves. Time windows of $\pm$3 transit duration were adopted for S18 (1800~s cadence), and $\pm$1.5 durations for S58 and S85 (120~s cadence).

\section{Light curve analysis}
\label{Section:lc}

\begin{table}%[h]
\renewcommand\arraystretch{1.5}
\centering
\caption{Priors and posterior constraints on transit parameters and orbital ephemeris derived from TESS light curve analysis. }
\label{tab:TESS_wlc}
\begin{tabular}{ccc}
\hline\hline
Parameter & Prior & Posterior Estimate \\
\hline
$i$ [$^{\circ}$] & $\mathcal{U}(80, 90)$ & $84.56^{+0.25}_{-0.30}$\\
$a/R_{\star}$ & $\mathcal{U}(5, 12.5)$ & $8.53^{+0.25}_{-0.29}$\\
$R_{\rm p}/R_{\star}$ & $\mathcal{U}(0.05, 0.15)$ & $0.08883^{+0.00091}_{-0.00078}$\\
$u_1$ & fixed & 1.64772 \\
$u_2$ & fixed & $-$3.16708 \\
$u_3$ & fixed & 3.76242 \\
$u_4$ & fixed & $-$1.48730 \\
\hline
$P$ [d] & -- & $4.73217441^{+0.00000055}_{-0.00000055}$ \\
$T_0$ [$\rm BJD_{TDB}$] & -- & $2458387.36996^{+0.00024}_{-0.00024}$ \\
\noalign{\medskip}
\hline
\end{tabular}
\end{table}

The light curves were modeled as a transit signal combined with correlated systematics. The transit component was computed with the \texttt{batman} Python package \citep{Kreidberg2015} assuming a circular orbit, with free parameters including: orbital period ($P$), orbital inclination ($i$), scaled semi-major axis ($a/R_{\star}$), planet-to-star radius ratio ($R_{\rm p}/R_{\star}$), mid-transit time ($T_{\rm mid}$), and limb-darkening coefficients (LDCs). Correlated systematics were modeled with Gaussian processes (GP; \citealp{Rasmussen2006, Gibson2012b}) applied to the full light curves, using the \texttt{george} \citep{Ambikasaran2015} and \texttt{celerite} \citep{Foreman-Mackey2017} Python packages. 

The TESS data, being broadband photometry, were treated as white light curves. Since the ground-based LBT/MODS and GTC/OSIRIS+ data are affected by strong systematics and intermittent gaps, limiting their ability to constrain the transit parameters independently, we first fitted the TESS light curves to determine the transit parameters and refine the ephemeris. These refined values were then fixed when fitting the MODS and OSIRIS+ light curves. 

\begin{figure}%[htb]
\centering
\includegraphics[width=\hsize]{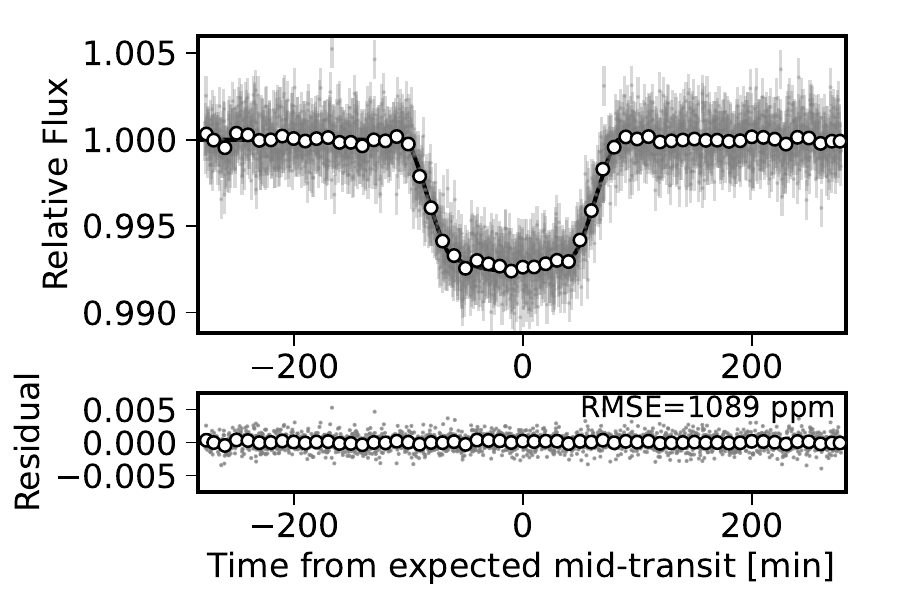}
\caption{Phase-folded TESS transit light curves of HAT-P-47. Top: detrended and normalized flux (gray circles), 10-min binned data (black circles), and the best-fit transit model (black line). Bottom: residuals between the data and model. }
\label{fig:TESS_folded}
\end{figure}

\subsection{TESS light curves}
\label{subsect:TESS-lc}
\subsubsection{Transit parameter fitting}
One-dimensional GP regression was implemented using the \texttt{celerite} Python package \citep{Foreman-Mackey2017}, with observation time serving as the input vector, which is well suited for the large number of data points in the TESS light curves. The GP mean function was constructed as the product of the transit model computed with \texttt{batman} and a linear baseline function of the form 
\begin{equation}
b(x,y)=c_0+c_1x+c_2y\;,
\end{equation}
where $x$ and $y$ denote the centroid coordinates of the target on the detector, and $c_0$, $c_1$, and $c_2$ are coefficients describing systematic trends. The covariance structure of the correlated noise was described by a Matérn-$3/2$ kernel, 
\begin{equation}
 k_{\rm M32}(\tau) = \sigma_k^2\left(1 + \frac{\sqrt{3}\tau}{\rho}\right) \exp\left( - \frac{\sqrt{3}\tau}{\rho}\right) + \sigma_w^2,
\end{equation}
where $\tau = \lvert t_i - t_j \rvert$ is the temporal separation between $i$-th and $j$-th measurements. The hyperparameters $\sigma_k$ and $\rho$ are the amplitude and characteristic timescale of the correlated noise, respectively, while $\sigma_w^2$ accounts for additional white-noise variance. 

All transit light curves were fitted jointly, sharing common transit parameters except for the mid-transit time $T_{\rm mid}$. The baseline coefficients and GP hyperparameters were allowed to vary independently for each transit to capture epoch-specific systematics. For the S18 light curves, the \texttt{batman} supersampling option was applied to correct for long-cadence smearing \citep{Kipping2010}.

We adopted a four-parameter nonlinear limb-darkening law, with coefficients ($u_1$, $u_2$, $u_3$, $u_4$) fixed to values derived from PHOENIX stellar atmosphere models \citep{Husser2013} using the Python tool developed by \citet{Espinoza2015}. The model grid was selected to match the stellar parameters of HAT-P-47 (effective temperature $T_{\rm eff}=6700$~K, surface gravity $\log g_{\star}=4.00$~(cgs), and metallicity ${\rm [Fe/H]}=0.0$). 

Model parameters were first optimized using a Nelder–Mead minimization to identify the maximum-likelihood solution, followed by Markov chain Monte Carlo (MCMC) sampling with the \texttt{emcee} Python package \citep{Foreman-Mackey2013} to estimate posterior distributions. The best-fit values and uncertainties were taken as the median and 16th–84th percentiles of the posterior samples. The MCMC analysis employed three ensembles of 240 walkers each, initialized near the optimal solution. After two burn-in phases of 5,000 steps, a production run of 50,000 steps was carried out with a thinning factor of 5. 

The derived transit parameters are listed in Table~\ref{tab:TESS_wlc}, and the phase-folded light curve is shown in Fig.~\ref{fig:TESS_folded}. The residuals have a scatter of 1089~ppm (root mean square error, RMSE). We obtain $i = 84.56^{+0.25}_{-0.30}\,^{\circ}$, $a/R_{\star} = 8.53^{+0.25}_{-0.29}$, and $R_{\rm p}/R_{\star} = 0.08883^{+0.00091}_{-0.00078}$, in agreement with the discovery values \citep[$i = 84.8 \pm 0.2^{\circ}$, $a/R_{\star} = 8.73\pm 0.20$, $R_{\rm p}/R_{\star} = 0.0890\pm 0.0013$;][]{Bakos2016} within $0.71\sigma$, $0.60\sigma$, and $0.13\sigma$, respectively.

\begin{figure}%[htb]
\centering
\includegraphics[width=\hsize]{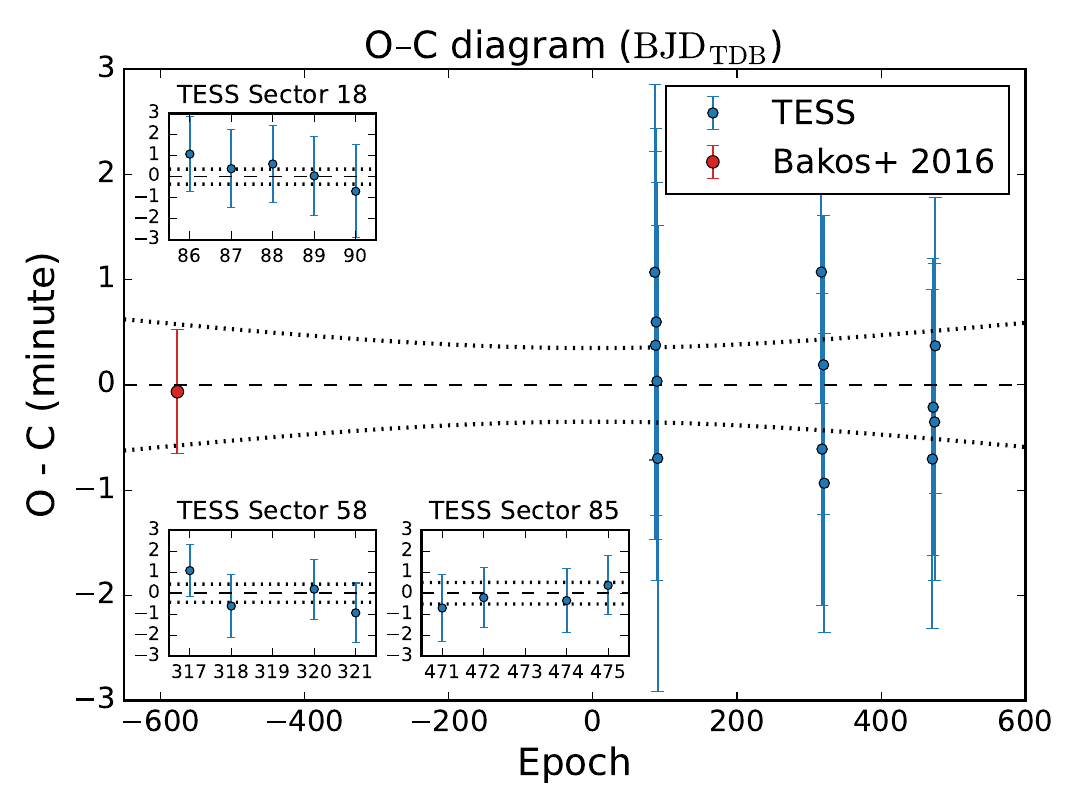}
\caption{Transit timing residuals of HAT-P-47 relative to a linear ephemeris. Points show the differences between the measured mid-transit times and model predictions. The dashed line marks zero residual, while the dotted lines indicate the $\pm 1\sigma$ uncertainties. The insets show zoom-ins for individual sectors. }
\label{fig:TESS_OC}
\end{figure}

\subsubsection{Orbital ephemeris refinement}
Mid-transit times were measured for each TESS event and combined with the value reported by \citet{Bakos2016} to refine the orbital ephemeris. All timestamps are expressed in Barycentric Julian Date in the Barycentric Dynamical Time standard ($\rm BJD_{TDB}$; \citealp{Eastman2010}). The mid-transit times $T_\mathrm{mid}$ were modeled as a function of epoch $E$ assuming a constant orbital period, adopting the linear ephemeris  
\begin{equation}
T_\mathrm{mid} = T_0 + PE,
\end{equation}
where $T_0$ denotes the reference mid-transit time at zero epoch, which was optimized to minimize the uncertainty of $T_0$. 

Figure~\ref{fig:TESS_OC} shows the timing residuals relative to the best-fit linear ephemeris, and the updated ephemeris is given in Table~\ref{tab:TESS_wlc}. We obtain a refined period of $P = 4.73217441 \pm 0.00000055$~d, consistent with the discovery value ($P = 4.732182 \pm 0.000013$~d) within $0.58\sigma$, with a $\sim$24-fold improvement in precision. The updated reference epoch is $T_0 = 2458387.36996 \pm 0.00024$ ($\rm BJD_{TDB}$).

\begin{table*}%[h]
\renewcommand\arraystretch{1.5}
\centering
\caption{Priors and posterior constraints on the parameters used in the white light curve analyses. }
\label{tab:wlc}
\resizebox{0.85\hsize}{!}{
\begin{tabular}{cccccc}
\hline\hline
Parameter & Prior & MODS1B Estimate & MODS1R Estimate & MODS2B Estimate & OSIRIS+ Estimate \\
\hline
$R_{\rm p}/R_{\star}$ & $\mathcal{U}(0.05,0.15)$ & $0.087176^{+0.003724}_{-0.003362}$ & $0.093953^{+0.006844}_{-0.005961}$ & $0.087176^{+0.003724}_{-0.003362}$ & $0.094830^{+0.005368}_{-0.006115}$ \\
$\Delta T_{\rm mid}$ [day]\tablefootmark{a} & $\mathcal{N}(0,\sigma^2)$\tablefootmark{a} & $0.000042^{+0.000139}_{-0.000146}$ & $0.000042^{+0.000139}_{-0.000146}$ & $0.000042^{+0.000139}_{-0.000146}$ & $-0.000244^{+0.000325}_{-0.000314}$ \\
$u_1$ & fixed & 1.08259 & 1.55650 & 1.07299 & 1.32449 \\
$u_2$ & fixed & $-2.00054$ & $-2.93371$ & $-1.98433$ & $-2.46316$ \\
$u_3$ & fixed & 3.17011 & 3.57819 & 3.16661 & 2.46316 \\
$u_4$ & fixed & $-1.37249$ & $-1.42371$ & $-1.37405$ & $-1.38624$ \\
$\sigma_w$~$[10^{-6}]$ & $\mathcal{U}(0.1, 10000)$ & $323.743^{+34.322}_{-37.487}$ & $348.859^{+28.993}_{-26.021}$ & $364.794^{+26.304}_{-24.006}$ &
$439.907^{+15.115}_{-14.894}$ \\
$\ln \sigma_k$ & $\mathcal{U}(-10,-1)$ & $-7.14858^{+0.44495}_{-0.44661}$ & $-6.55471^{+0.40996}_{-0.34463}$ & $-6.49073^{+0.43026}_{-0.58190}$ & $-6.72040^{+0.32994}_{-0.27327}$ \\
$\ln \tau_t$ & $\mathcal{U}(-6,5)$ & $-2.09683^{+0.69484}_{-0.92273}$ & $-1.96482^{+0.52079}_{-0.47389}$ & $-1.8899^{+0.53193}_{-0.64072}$ & $-2.05794^{+3.36071}_{-0.47960}$ \\
$\ln \tau_x$ & $\mathcal{U}(-5,5)$ & $2.87700^{+1.40008}_{-1.50351}$ & $4.03296^{+0.66101}_{-1.51146}$ & $3.78833^{+0.74406}_{-0.76344}$ & $4.43325^{+0.40956}_{-0.61319}$ \\
$\ln \tau_y$ & $\mathcal{U}(-5,5)$ & $2.93534^{+1.25875}_{-0.81441}$ & $3.44580^{+0.75596}_{-0.66389}$ & $3.79417^{+0.68328}_{-0.80957}$ & $3.93137^{+0.73286}_{-0.92949}$ \\
$\ln \tau_{s_y}$ & $\mathcal{U}(-5,5)$ & $0.78427^{+0.50629}_{-0.40774}$ & $1.53058^{+0.38014}_{-0.34338}$ & $2.93465^{+0.54835}_{-0.75702}$ & $4.19516^{+0.49328}_{-0.55279}$ \\
$\ln \tau_\theta$ & $\mathcal{U}(-5,5)$ & $4.44371^{+0.43037}_{-1.44408}$ & $4.35523^{+0.46210}_{-0.59754}$ & $4.64115^{+0.26105}_{-0.56667}$ & $3.53354^{+0.89737}_{-0.64912}$ \\
\noalign{\medskip}
\hline
\end{tabular}}
\tablefoot{
\tablefoottext{a}{For MODS: $\Delta T_{\rm mid} = {\rm BJD_{TDB}} - 2458434.69250$, $\sigma = 0.00024$. For OSIRIS+: $\Delta T_{\rm mid} = {\rm BJD_{TDB}} -2460275.50835$, $\sigma = 0.00033$. }
}
\end{table*}

\begin{figure*}%[htb]
\centering
\includegraphics[width=0.75\hsize]{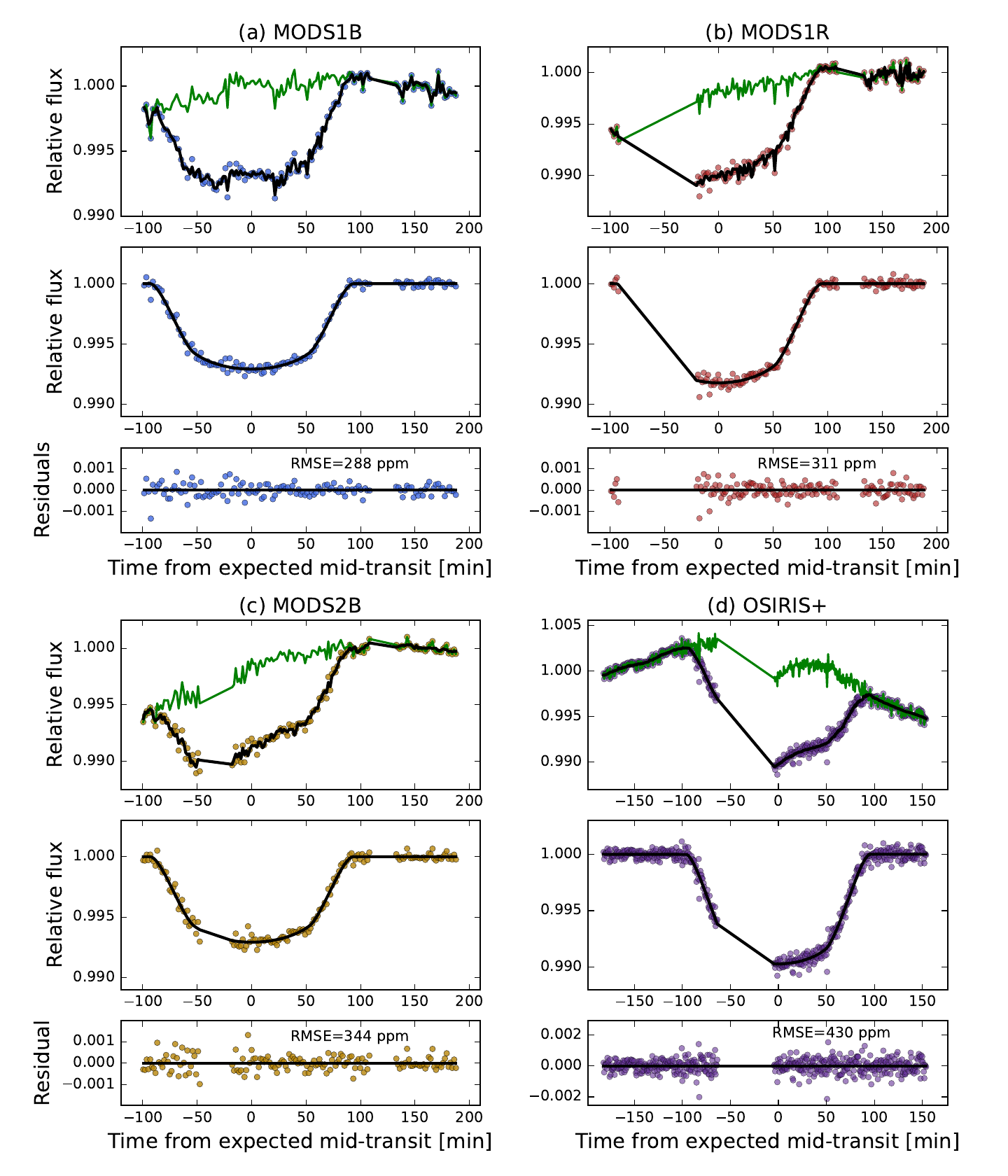}
\caption{White light curves of HAT-P-47 from LBT/MODS (panels a-c) and GTC/OSIRIS+ (panel d). From top to bottom in each panel: raw light curves (colored circles) with best-fit model (black line) and GP systematics (green line), detrended light curves, and residuals. }
\label{fig:wlc}
\end{figure*}

\subsection{LBT/MODS light curves}
\label{subsect:MODS-lc}
\subsubsection{White light curves}
Correlated noise in both white and spectroscopic light curves was modeled with GP using the \texttt{george} package \citep{Ambikasaran2015}. The GP mean function consisted solely of the analytic transit model computed with \texttt{batman}, with no additional parametric baseline. Systematics were fully described by the GP covariance function. We adopted a multidimensional Matérn $3/2$ kernel expressed as the product of independent kernels,  
\begin{equation}
 k_{\rm total}(\mathbf{x}_i, \mathbf{x}_j) = \sigma_k^2 \prod_{\alpha} k_{\rm M32}(r_{\alpha})\;,
\end{equation}
where $\sigma_k$ is the amplitude and $r_{\alpha} = |\mathbf{x}_{\alpha,i}-\mathbf{x}_{\alpha,j}|/\tau_{\alpha}$, with ${\tau_{\alpha}}$ the characteristic length scale of each input variable. The Matérn $3/2$ kernel is defined as
\begin{equation}
 k_{\rm M32}(r) = \left(1 + \sqrt{3} r \right) \exp\left( - \sqrt{3} r \right).
 \label{eq:M32}
\end{equation}
The input vector is $\mathbf{x} = (t, x, y, s_y, \theta)$, where $t$ is time, $x$ and $y$ are spatial and spectral drifts, $s_y$ is the full width at half maximum (FWHM) of the stellar point spread function (PSF), and $\theta$ is the instrumental rotation angle. All inputs were standardized prior to modeling. An additional white-noise term $\sigma_w$ was included by rescaling the photometric uncertainties. 

The MODS1B, MODS1R, and MODS2B white light curves were fitted jointly. The transit parameters $P$, $i$, and $a/R_{\star}$ were fixed to the values derived in Section~\ref{subsect:TESS-lc} (Table~\ref{tab:TESS_wlc}), and the LDCs were held at precomputed values. A common $R_{\rm p}/R_{\star}$ was adopted for the overlapping wavelength range of MODS1B and MODS2B, while an independent value was fitted for MODS1R. A shared mid-transit time $T_{\rm mid}$ was fitted across all channels, with a Gaussian prior centered on the predicted ephemeris. The GP hyperparameters \{$\sigma_w$, $\sigma_k$, $\tau_t$, $\tau_x$, $\tau_y$, $\tau_{s_y}$, $\tau_\theta$\} were fitted independently for each channel to account for instrument-specific systematics. Posterior sampling was performed using three MCMC ensembles of 64 walkers. After two burn-in phases of 1,000 steps, a production run of 5,000 steps was carried out. 

Figure~\ref{fig:wlc} shows the white light curves with the best-fit transit models and GP systematics components. The RMSE is 288, 311, and 344~ppm for MODS1B, MODS1R, and MODS2B, corresponding to 2.8, 4.8, and 3.7 times photon noise, respectively. The adopted priors and posterior constraints are listed in Table~\ref{tab:wlc}. 

\begin{figure*}%[htb]
\centering
\includegraphics[width=0.75\hsize]{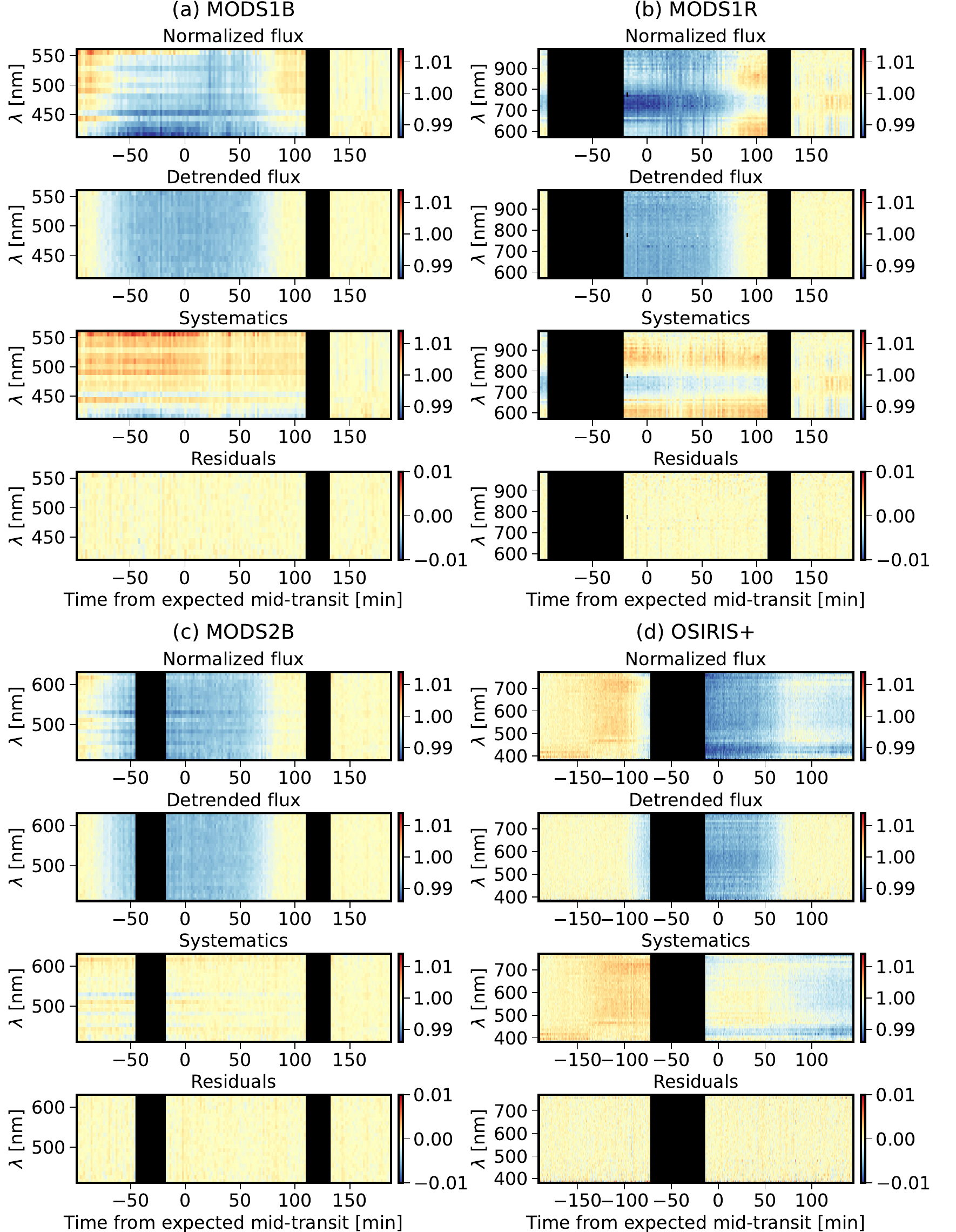}
\caption{Spectroscopic light curves of HAT-P-47 from LBT/MODS (panels a-c) and GTC/OSIRIS+ (panel d). From top to bottom in each panel: matrices of raw light curves, detrended light curves, GP systematics, and residuals. Black boxes indicate time intervals with no available data in the corresponding wavelength channels. }
\label{fig:slc_2D}
\end{figure*}

\subsubsection{Spectroscopic light curves}
We adopted a direct-fitting approach for the spectroscopic light curves, in which each wavelength channel was modeled independently rather than applying a divide-white method \citep{Gibson2013} to remove common-mode systematics derived from the white-light curve. A previous benchmark study on transiting white dwarfs has shown that divide-white can underestimate uncertainties in transit depth and introduce artificial spectral features \citep{Jiang2022}. We compared both approaches and found that divide-white yields larger uncertainties in transit depths, while the resulting transmission spectra are consistent within the uncertainties. This indicates that a common-mode correction does not adequately capture the systematics. The presence of wavelength-dependent systematics in spectroscopic light curves (Fig.~\ref{fig:slc_2D}) further supports this. We therefore adopted the direct-fitting approach. 

Each spectroscopic light curve was fitted independently for $R_{\rm p}/R_{\star}$. The transit parameters $P$, $i$ and $a/R_{\star}$ were fixed to the TESS-derived values (Table~\ref{tab:TESS_wlc}), while $T_{\rm mid}$ was fixed to the median values from the joint white light-curves fit (Table~\ref{tab:wlc}). LDCs were fixed to precomputed values. The GP covariance matrix included five input vectors $\{t, x, y, s_y, \theta\}$, with corresponding length scales and hyperparameters ($\sigma_k$ and $\sigma_w$) fitted simultaneously. Priors were identical to those adopted in the white-light-curve analysis. For the MODS1R 766--776~nm and 776--786~nm passbands, a single outlier was removed prior to fitting. 

Posterior sampling was performed using three MCMC ensembles of 32 walkers, with two burn-in phases of 1,000 steps followed by a 5,000-step production run. Figure~\ref{fig:slc_2D} shows the matrix of spectroscopic light curves, together with the best-fit systematics models and residuals. 

\subsection{GTC/OSIRIS+ light curves}
\label{subsect:OSIRIS-lc}
\subsubsection{White light curve}
The OSIRIS+ white light curve was fitted independently using the same GP framework implemented with \texttt{george}. The GP mean function consisted solely of the \texttt{batman} transit model, with no additional parametric baseline. Systematics were fully captured by the GP covariance. The GP inputs were $\{t, x, y, s_y, \theta\}$. The transit parameters $P$, $i$, and $a/R_{\star}$ were fixed to the TESS-derived values (Table~\ref{tab:TESS_wlc}), while $T_{\rm mid}$ was assigned a Gaussian prior centered on the prediction from the refined ephemeris. LDCs were fixed to precomputed values. Posterior sampling used three MCMC ensembles of 32 walkers, with two burn-in phases of 1,000 steps followed by a 5,000-step production run. 

The white light curve, best-fit model, and GP systematics are shown in Fig.~\ref{fig:wlc}, and the fitted parameters are listed in Table~\ref{tab:wlc}. The residual scatter has an RMSE of 430~ppm, corresponding to 4.4 times photon noise. 

\subsubsection{Spectroscopic light curves}
The OSIRIS+ spectroscopic light curves were analyzed using the same direct-fitting approach as for the MODS data, without applying the divide-white method. The divide-white procedure yields highly uneven uncertainties in transit depths across wavelength channels, far exceeding expectations from photon noise, indicating that a wavelength-independent common mode does not adequately describe the systematics. We therefore adopted the direct-fitting approach. 

For each spectroscopic channel, $R_{\rm p}/R_{\star}$ was fitted independently. The transit parameters $P$, $i$, and $a/R_{\star}$ were fixed to the TESS-derived values, while $T_{\rm mid}$ was fixed to the median from the white-light fit. LDCs were fixed to the precomputed values. The GP covariance included $\{t, x, y, s_y, \theta\}$ with corresponding length scales and hyperparameters ($\sigma_k$ and $\sigma_w$) fitted simultaneously. Priors were identical to those used for the white-light analysis, except for $\tau_t$, whose log-uniform prior was restricted to $[-6, 1]$, instead of $[-6, 5]$, to avoid multimodal solutions encountered in some of the wavelength channels. 

Posterior sampling used three MCMC ensembles with 32 walkers each, with two burn-in phases of 1,000 steps followed by a 5,000-step production run. The spectroscopic light curves and best-fit models are shown in Fig.~\ref{fig:slc_2D}.

\begin{figure}%[htb]
\centering
\includegraphics[width=\hsize]{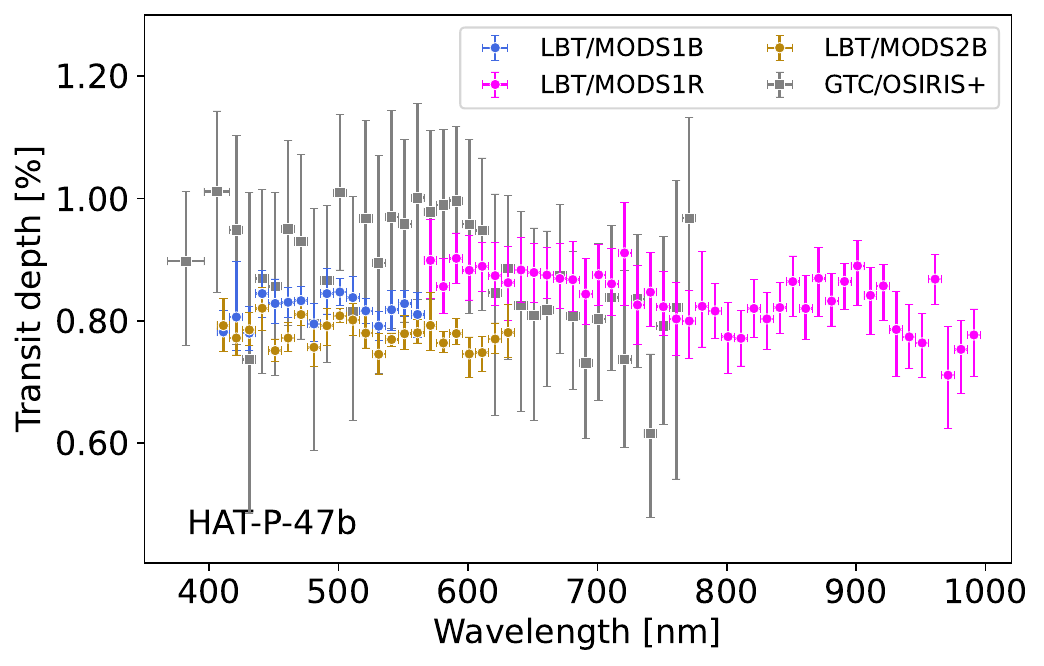}
\caption{Individual transmission spectra of HAT-P-47b derived from LBT/MODS1B (blue), LBT/MODS1R (magenta), LBT/MODS2B (yellow), and GTC/OSIRIS+ (gray). }
\label{fig:MODS_OSIRIS_transpec}
\end{figure}

\section{Transmission spectra and spectral retrieval}
\label{Section:retrieval}
\subsection{Transmission spectra}
The resulting transmission spectra from MODS and OSIRIS+ are shown in Fig.~\ref{fig:MODS_OSIRIS_transpec} and listed in Table~\ref{tab:spectra}. Significant offsets are observed between MODS1 and MODS2 in their overlapping wavelengths (16 passbands over 406--566~nm for MODS1B-MODS2B and 7 passbands over 566--636~nm for MODS1R-MODS2B). 

To investigate the consistency between different MODS channels, we performed a diagnostic test using the common wavelength regions. For each pair of overlapping channels, we constructed pseudo white light curves restricted to the common wavelength range and derived a joint transit depth. This reference value was then compared to those obtained from independent fits to each dataset, providing an estimate of their relative baseline offsets. Applying these offsets to the transmission spectra significantly improves the agreement in the overlapping regions, reducing $\chi^2$ from 59.16 to 7.44 for 23 degrees of freedom (dof). This demonstrates that the apparent discrepancies are primarily due to channel-dependent baseline differences rather than intrinsic inconsistencies between the datasets. 

We stress that this correction is used solely as a consistency check and is not applied to the final transmission spectra. Instead, in the subsequent analysis, these baseline differences are accounted for by introducing wavelength-independent offset parameters for each dataset ($\Delta_\mathrm{offset,MODS1B}$ and $\Delta_\mathrm{offset,MODS2B}$) relative to MODS1R, thereby properly propagating the associated uncertainties. A similar offset parameter ($\Delta_\mathrm{offset,OSIRIS+}$) is included for the OSIRIS+ dataset. To account for the relative weighting of MODS and OSIRIS+ data, given their different levels of correlated noise, we also introduce dataset-dependent error-scaling factors ($f_{\sigma,\rm MODS}$ and $f_{\sigma,\rm OSIRIS+}$) that rescale the reported uncertainties. 

We first compared the observed transmission spectra of HAT-P-47b with two simple parametric models: a wavelength-independent flat model and a linear trend in $(\ln \lambda,\; R_{\rm p}/R_{\star})$ space. If the atmospheric opacity follows a power-law cross section, $\sigma(\lambda) = \sigma_0 \left( \frac{\lambda}{\lambda_0} \right)^{\alpha}$, a slope is expected in the transmission spectrum \citep{LecavelierDesEtangs2008}. The corresponding variation of the apparent planetary radius with wavelength is
\begin{equation}
\frac{d R_{\rm p}}{d \ln \lambda} = \alpha H = \alpha \frac{k_{\rm B} T}{\mu g_{\rm p}}\;,
\end{equation}
where $H$ is the atmospheric scale height, $k_{\rm B}$ the Boltzmann constant, $T$ the atmospheric temperature, $\mu$ the mean molecular weight, and $g_{\rm p}$ the planetary surface gravity. For Rayleigh scattering by molecular hydrogen, $\alpha = -4$ is expected. 

Parameter estimation and model comparison were performed using the nested sampling algorithm implemented in the \texttt{PyMultiNest} Python package \citep{Buchner2014}, adopting 250 live points. Model comparison is based on the Bayesian evidence $\mathcal{Z}$. Differences in log-evidence ($\Delta \ln \mathcal{Z}$) are interpreted following \citet{Trotta2008} criteria: $\Delta \ln \mathcal{Z} \geq 5.0$ indicates strong evidence, $2.5 \leq \Delta \ln \mathcal{Z} < 5.0$ moderate evidence, $1.0 \leq \Delta \ln \mathcal{Z} < 2.5$ weak evidence, and $\Delta \ln \mathcal{Z} < 1.0$ inconclusive. The numerical uncertainty in $\ln \mathcal{Z}$ is estimated to be $\sim 0.1$--0.4, depending on model complexity. 

For the MODS data alone, the sloped model is favored over the flat model, with $\ln \mathcal{Z} = 529.77$ versus 524.20 ($\Delta \ln \mathcal{Z} = 5.58$), indicating strong evidence for a wavelength-dependent trend. Adopting $T = T_{\rm eq} = 1605$~K, $\log g_{\rm p} = 2.47$~(cgs), $R_{\star} = 1.515$~$R_{\odot}$, and $\mu = 2.3$~$m_{\rm H}$, we obtain $\alpha = -2.73 \pm 0.69$. For the OSIRIS+ dataset, the sloped model is similarly preferred ($\Delta \ln \mathcal{Z} = 3.68$), yielding $\alpha = -6.48 \pm 2.06$. A joint fit to the MODS and OSIRIS+ data further strengthens the preference for a sloped model ($\Delta \ln \mathcal{Z} = 8.48$), with $\alpha = -3.14 \pm 0.67$. 

Within the uncertainties, the three independent estimates of $\alpha$ are mutually consistent, indicating that both datasets support the presence of a wavelength-dependent slope in the optical transmission spectrum. The value derived from the joint fit is consistent with the Rayleigh-scattering expectation within $1.3\sigma$. We note that the inferred slope is degenerate with the channel-dependent offset parameters, which may influence the derived value of $\alpha$. This simple parametric slope model is intended only as a phenomenological description of wavelength-dependent trends in the data and is not directly tied to the scattering prescription adopted in the atmospheric retrieval analysis. 

\subsection{Atmospheric retrieval analysis}
\subsubsection{Model setup}
We performed Bayesian retrieval analyses to constrain the atmospheric properties of HAT-P-47b using its optical transmission spectra. The MODS and OSIRIS+ datasets were first analyzed independently and then combined in a joint retrieval. Forward atmospheric models were generated with the Python package \texttt{petitRADTRANS} \citep{Molliere2019}, assuming a one-dimensional atmosphere in hydrostatic equilibrium with spherical transmission geometry. 

Owing to the moderate spectral resolution and limited wavelength coverage, the temperature structure was assumed to be isothermal, parameterized by a single free parameter $T_{\rm iso}$. The pressure grid spanned $10^{2}$--$10^{-6}$~bar and was discretized into 100 layers uniformly spaced in logarithmic pressure. The reference pressure $P_0$ was treated as a free parameter and was defined the pressure at the planetary radius, fixed to $R_{\rm p} = 1.310$~$R_{\rm Jup}$ based on the refined TESS $R_{\rm p}/R_{\star}$. 

The atmosphere was assumed to be dominated by $\rm H_2$ and $\rm He$ in a 3:1 mass ratio. Opacity sources included 15 atomic and molecular absorbers ($\rm Na$, $\rm K$, $\rm TiO$, $\rm VO$, $\rm H_2O$, $\rm CH_4$, $\rm CO$, $\rm CO_2$, $\rm HCN$, $\rm C_2H_2$, $\rm PH_3$, $\rm NH_3$, $\rm H_2S$, $\rm SiO$, and $\rm FeH$), as well as collision-induced absorption (CIA) from $\rm H_2$–$\rm H_2$ and $\rm H_2$–He pairs \citep[and the references therein]{Borysow1988,Borysow1989a,Borysow1989b,Borysow2001,Borysow2002,Richard2012}, and Rayleigh scattering by $\rm H_2$ \citep{Dalgarno1962} and He \citep{Chan1965}. Opacities were computed using the pre-computed correlated-$k$ opacity tables of \texttt{petitRADTRANS 2.6.7} at a resolving power of $R \sim 1000$ \citep{Molliere2019}. The adopted atomic and molecular line-list references follow those listed in Table~2 of \citet{Molliere2019}.

We explored two alternative chemical parameterizations. In the equilibrium chemistry scenario, molecular abundances were interpolated from thermochemical grids and parameterized by the atmospheric metallicity ($\log Z$) and carbon-to-oxygen ratio (C/O). In the free chemistry scenario, the mass fractions ($X_i$) of the individual opacity species were treated as independent free parameters. 

Clouds were described using a patchy terminator model following \citet{MacDonald2017}, in which a fraction $\phi$ of the planetary limb was covered by an opaque gray cloud deck, while the remainder was cloud-free. The cloudy sector was characterized by a cloud-top pressure $P_{\rm cloud}$, and an additional scattering enhancement factor $A_{\rm RS}$ was included to capture potential Rayleigh-like slopes.

Following our previous work \citep{Wang2026}, we also explored the impact of the transit light source effect \citep{Rackham2018,Rackham2019}, despite HAT-P-47 being a relatively inactive F star with a chromospheric activity index of $\log R^\prime_{\rm HK}=-5.125 \pm 0.015$ \citep[][although a value of $-4.78 \pm 0.10$ has also been reported by \citealt{Claudi2024}]{Bakos2016}. For this purpose, we performed additional retrievals on the combined transmission spectra using a hybrid model that integrates a free-chemistry atmosphere with stellar contamination from unocculted spots and faculae. The wavelength-dependent correction to the transit depth is expressed as
\begin{equation}
 \begin{split}
 D_{\lambda, \rm c} &= D_{\lambda} S(\lambda, T_{\rm phot}) \; / \; \left[ f_{\rm spot}S(\lambda, T_{\rm spot}) \right. \\ &\left. + f_{\rm facu}S(\lambda, T_{\rm facu}) + (1 - f_{\rm spot} - f_{\rm facu})S(\lambda, T_{\rm phot}) \right]\;,
 \label{eq:TDc}
 \end{split}
\end{equation}
where $D_{\lambda}$ and $D_{\lambda, \rm c}$ denote the modeled transit depths without and with stellar contamination, respectively. The parameters $T_{\rm phot}$, $T_{\rm spot}$, and $T_{\rm facu}$ correspond to the temperatures of the stellar photosphere, spots, and faculae, while $1 - f_{\rm spot} - f_{\rm facu}$, $f_{\rm spot}$ and $f_{\rm facu}$ represent their respective covering fractions. The function $S(\lambda, T)$ denotes the stellar spectrum interpolated from the PHOENIX spectral library \citep{Husser2013}. 

\begin{table*}%[htb]
\renewcommand\arraystretch{1.5} 
\caption[]{Priors and derived posteriors for the parameters used in the spectral retrieval analyses.}
\begin{center}
\scalebox{0.8}{
\begin{tabular}{cccccccc}
\hline\hline
\multirow{2}{*}{Parameter} & \multirow{2}{*}{Prior}  & \multicolumn{2}{c}{LBT/MODS Posterior} & \multicolumn{2}{c}{GTC/OSIRIS+ Posterior} & \multicolumn{2}{c}{Joint Posterior} \\
\cmidrule(lr){3-4}
\cmidrule(lr){5-6}
\cmidrule(lr){7-8}
 &   & Equilibrium Chem. & Free Chem. & Equilibrium Chem. & Free Chem. & Equilibrium Chem. & Free Chem. \\
\hline
$T_\mathrm{iso}$ [K] 
& ${\mathcal{U}}(1000,2000)$ 
& $1539^{+144}_{-186}$ 
& $1446^{+270}_{-239}$ 
& $1358^{+504}_{-197}$
& $1605^{+259}_{-361}$ 
& $1618^{+170}_{-142}$ 
& $1669^{+202}_{-261}$ \\
$\log P_0$ [$\log {\rm bar}$]
& $\mathcal{U}(-6,2)$ 
& $-3.49^{+0.69}_{-0.33}$
& $-3.98^{+0.46}_{-0.35}$ 
& $-0.91^{+0.61}_{-0.81}$
& $-2.76^{+0.90}_{-0.81}$ 
& $-3.42^{+0.51}_{-0.37}$ 
& $-3.98^{+0.45}_{-0.31}$ \\
$\log P_\mathrm{cloud}$ [$\log {\rm bar}$] 
& $\mathcal{U}(-6,2)$ 
& $-1.49^{+2.19}_{-1.71}$ 
& $-1.01^{+1.85}_{-1.90}$ 
& $-0.58^{+1.50}_{-1.97}$
& $-0.81^{+1.74}_{-1.86}$ 
& $-1.11^{+2.03}_{-2.06}$ 
& $-0.99^{+1.90}_{-1.96}$ \\
$\log A_\mathrm{RS}$ 
& $\mathcal{U}(-4,4)$ 
& $3.65^{+0.25}_{-0.62}$ 
& $3.61^{+0.27}_{-0.45}$ 
& $0.98^{+1.70}_{-1.73}$
& $2.29^{+0.98}_{-1.18}$ 
& $3.63^{+0.27}_{-0.51}$ 
& $3.65^{+0.24}_{-0.42}$\\
$\phi$ 
& $\mathcal{U}(0,1)$ 
& $0.82^{+0.08}_{-0.09}$ 
& $0.80^{+0.10}_{-0.09}$
& $0.39^{+0.30}_{-0.23}$
& $0.70^{+0.18}_{-0.24}$ 
& $0.78^{+0.09}_{-0.09}$ 
& $0.78^{+0.09}_{-0.09}$ \\
$\mathrm{C/O}$ 
& $\mathcal{U}(0.1,1.6)$ 
& $0.65^{+0.46}_{-0.34}$ 
& --
& $1.18^{+0.27}_{-0.42}$ 
& -- 
& $0.62^{+0.37}_{-0.35}$ 
& -- \\
$\log Z$ [$\log Z_\odot$]
& $\mathcal{U}(-2,3)$ 
& $-1.11^{+1.12}_{-0.53}$ 
& -- 
& $-0.83^{+0.94}_{-0.69}$ 
& -- 
& $-1.29^{+1.23}_{-0.47}$ 
& -- \\
$\log X_\mathrm{Na}$ 
& $\mathcal{U}(-12,0)$  
& -- 
& $-6.97^{+2.82}_{-2.96}$
& -- 
& $-3.72^{+1.99}_{-3.41}$
& -- 
& $-7.14^{+3.23}_{-3.15}$ \\
$\log X_\mathrm{K}$ 
& $\mathcal{U}(-12,0)$  
& -- 
& $-7.99^{+2.46}_{-2.20}$ 
& -- 
& $-7.18^{+2.61}_{-2.62}$
& -- 
& $-8.07^{+2.62}_{-2.42}$ \\
$\log X_\mathrm{TiO}$ 
& $\mathcal{U}(-12,-3)$ \tablefootmark{a} 
& -- 
& $-6.81^{+0.69}_{-0.65}$ 
& -- 
& $-8.94^{+2.27}_{-1.73}$
& -- 
& $-6.86^{+0.64}_{-0.63}$ \\
$\log X_\mathrm{VO}$ 
& $\mathcal{U}(-12,-3)$ \tablefootmark{a}
& -- 
& $-8.73^{+1.78}_{-1.86}$ 
& -- 
& $-9.89^{+1.56}_{-1.34}$
& -- 
& $-9.24^{+1.72}_{-1.71}$ \\
$\log X_\mathrm{H_2O}$ 
& $\mathcal{U}(-12,0)$ 
& -- 
& $-6.50^{+2.81}_{-3.11}$ 
& -- 
& $-6.81^{+3.14}_{-3.11}$ 
& -- 
& $-6.90^{+2.94}_{-3.03}$ \\
$\log X_\mathrm{CH_4}$ 
& $\mathcal{U}(-12,0)$ 
& -- 
& $-5.93^{+3.32}_{-3.42}$ 
& -- 
& $-6.23^{+3.30}_{-3.34}$ 
& -- 
& $-6.11^{+3.46}_{-3.79}$ \\
$\log X_\mathrm{CO}$ 
& $\mathcal{U}(-12,0)$ 
& -- 
& $-6.10^{+3.37}_{-3.28}$
& -- 
& $-6.27^{+3.22}_{-3.38}$
& -- 
& $-6.09^{+3.13}_{-3.68}$ \\
$\log X_\mathrm{CO_2}$ 
& $\mathcal{U}(-12,0)$ 
& -- 
& $-5.93^{+3.30}_{-3.69}$ 
& -- 
& $-6.16^{+3.31}_{-3.59}$
& -- 
& $-6.46^{+3.48}_{-3.44}$ \\
$\log X_\mathrm{HCN}$ 
& $\mathcal{U}(-12,0)$ 
& -- 
& $-6.76^{+3.50}_{-3.04}$ 
& -- 
& $-6.50^{+3.40}_{-3.34}$ 
& -- 
& $-6.83^{+3.36}_{-3.38}$ \\
$\log X_\mathrm{C_2H_2}$ 
& $\mathcal{U}(-12,0)$ 
& -- 
& $-5.98^{+3.30}_{-3.72}$ 
& -- 
& $-6.52^{+3.31}_{-3.34}$ 
& -- 
& $-6.42^{+3.38}_{-3.53}$ \\
$\log X_\mathrm{PH_3}$ 
& $\mathcal{U}(-12,0)$ 
& -- 
& $-6.49^{+3.56}_{-3.28}$ 
& -- 
& $-6.32^{+3.46}_{-3.44}$
& -- 
& $-6.42^{+3.56}_{-3.49}$ \\
$\log X_\mathrm{NH_3}$ 
& $\mathcal{U}(-12,0)$ 
& -- 
& $-6.72^{+3.13}_{-3.19}$ 
& -- 
& $-6.53^{+3.02}_{-3.08}$
& -- 
& $-6.66^{+3.08}_{-3.36}$ \\
$\log X_\mathrm{H_2S}$ 
& $\mathcal{U}(-12,0)$ 
& -- 
& $-6.08^{+3.27}_{-3.42}$ 
& -- 
& $-6.70^{+3.41}_{-3.38}$
& -- 
& $-6.46^{+3.37}_{-3.55}$ \\
$\log X_\mathrm{SiO}$ 
& $\mathcal{U}(-12,0)$ 
& -- 
& $-6.48^{+3.32}_{-3.46}$ 
& -- 
& $-6.06^{+3.15}_{-3.30}$
& -- 
& $-6.28^{+3.34}_{-3.43}$ \\
$\log X_\mathrm{FeH}$ 
& $\mathcal{U}(-12,0)$ 
& -- 
& $-7.80^{+2.21}_{-2.33}$ 
& -- 
& $-8.09^{+2.51}_{-2.50}$
& -- 
& $-7.67^{+2.20}_{-2.72}$ \\
$\Delta_\mathrm{offset,MODS1B}$ [ppm]
& $\mathcal{U}(-5000,5000)$ 
& $675^{+108}_{-91}$ 
& $591^{+120}_{-93}$ 
& -- 
& -- 
& $716^{+105}_{-107}$ 
& $660^{+117}_{-112}$ \\
$\Delta_\mathrm{offset,MODS2B}$ [ppm]
& $\mathcal{U}(-5000,5000)$ 
& $1064^{+109}_{-86}$ 
& $980^{+106}_{-93}$ 
& -- 
& -- 
& $1102^{+152}_{-140}$ 
& $1050^{+104}_{-98}$ \\
$\Delta_\mathrm{offset,OSIRIS+}$ [ppm]
& $\mathcal{U}(-5000,5000)$ 
& --
& -- 
& -- 
& -- 
& $-15^{+152}_{-140}$ 
& $-65^{+148}_{-141}$ \\
$f_{\rm \sigma, MODS}$
& $\mathcal{U}(0.1,10)$ 
& $0.74^{+0.06}_{-0.05}$
& $0.74^{+0.06}_{-0.05}$ 
& -- 
& --  
& $0.74^{+0.06}_{-0.06}$
& $0.73^{+0.06}_{-0.05}$ \\
$f_{\rm \sigma, OSIRIS+}$
& $\mathcal{U}(0.1,10)$ 
& --
& --
& $0.58^{+0.08}_{-0.06}$
& $0.59^{+0.08}_{-0.06}$ 
& $0.59^{+0.07}_{-0.06}$
& $0.60^{+0.06}_{-0.06}$ \\
\hline
\end{tabular}
}
\tablefoot{
\tablefoottext{a}{The prior ranges for TiO and VO were restricted to $[-12, -3]$, instead of $[-12, 0]$ adopted for the other species, in order to avoid multimodal solutions with abundances exceeding those predicted by equilibrium chemistry. This range remains broader than the abundances expected across the explored C/O and $\log Z$ parameter space and does not affect the inferred atmospheric properties. }
}
\label{tab:retrieved_param}  
\end{center}
\end{table*}

\begin{table}%[htb]
\renewcommand\arraystretch{1.5} 
\caption[]{Summary of Bayesian spectral retrieval statistics. }
\begin{center}
\scalebox{0.8}{
\begin{tabular}{cccccccc}
\hline\hline
\multirow{2}{*}{\#} & \multirow{2}{*}{Model} & \multicolumn{2}{c}{LBT/MODS} & \multicolumn{2}{c}{GTC/OSIRIS+} & \multicolumn{2}{c}{Joint} \\
\cmidrule(lr){3-4}
\cmidrule(lr){5-6}
\cmidrule(lr){7-8}
 & & $\ln\mathcal{Z}$ & $\Delta\ln\mathcal{Z}$ & $\ln\mathcal{Z}$ & $\Delta\ln\mathcal{Z}$ & $\ln\mathcal{Z}$ & $\Delta\ln\mathcal{Z}$ \\\noalign{\smallskip}
\hline\noalign{\smallskip}
\multicolumn{8}{l}{A. Simple assumption} \\\noalign{\smallskip}
{\it i} & Flat line 
& 524.20 & $-$5.58 
& 202.84 & $-$3.68 
& 728.22 & $-$8.48 \\
{\it ii} & Sloped line 
& 529.77 & 0
& 206.52 & 0 
& 736.70 & 0 \\
\noalign{\smallskip}
\hline\noalign{\smallskip}
\multicolumn{8}{l}{B. Atmospheric retrieval assuming equilibrium chemistry} \\\noalign{\smallskip}
{\it i} & Full model 
& 531.18 & 0 
& 207.40 & 0 
& 741.09 & 0 \\
\noalign{\smallskip}
\hline\noalign{\smallskip}
\multicolumn{8}{l}{C. Atmospheric retrieval assuming free chemistry} \\\noalign{\smallskip}
{\it i} & Full model 
& 531.46 & 0
& 206.02 & 0 
& 740.32 & 0 \\
{\it ii} & No Na 
& 531.06 & $-$0.40 
& 205.14 & $-$0.88 
& 740.14 & $-$0.18 \\
{\it iii} & No K 
& 531.47 & 0.01 
& 205.89 & $-$0.13
& 740.28 & $-$0.04 \\
{\it iv} & No TiO 
& 528.78 & $-$2.68  
& 205.68 & $-$0.34 
& 736.86 & $-$3.44 \\
{\it v} & No VO 
& 531.68 & 0.22 
& 206.37 & 0.35 
& 741.31 & 0.99 \\
{\it vi} & No $\rm H_2O$ 
& 530.89 & $-$0.57 
& 205.98 & $-$0.04 
& 740.44 & 0.12 \\
\noalign{\smallskip}
\hline
\end{tabular}
}
\label{tab:model_cmp}  
\end{center}
\end{table}

\begin{figure*}%[htb]
\centering
\includegraphics[width=0.85\hsize]{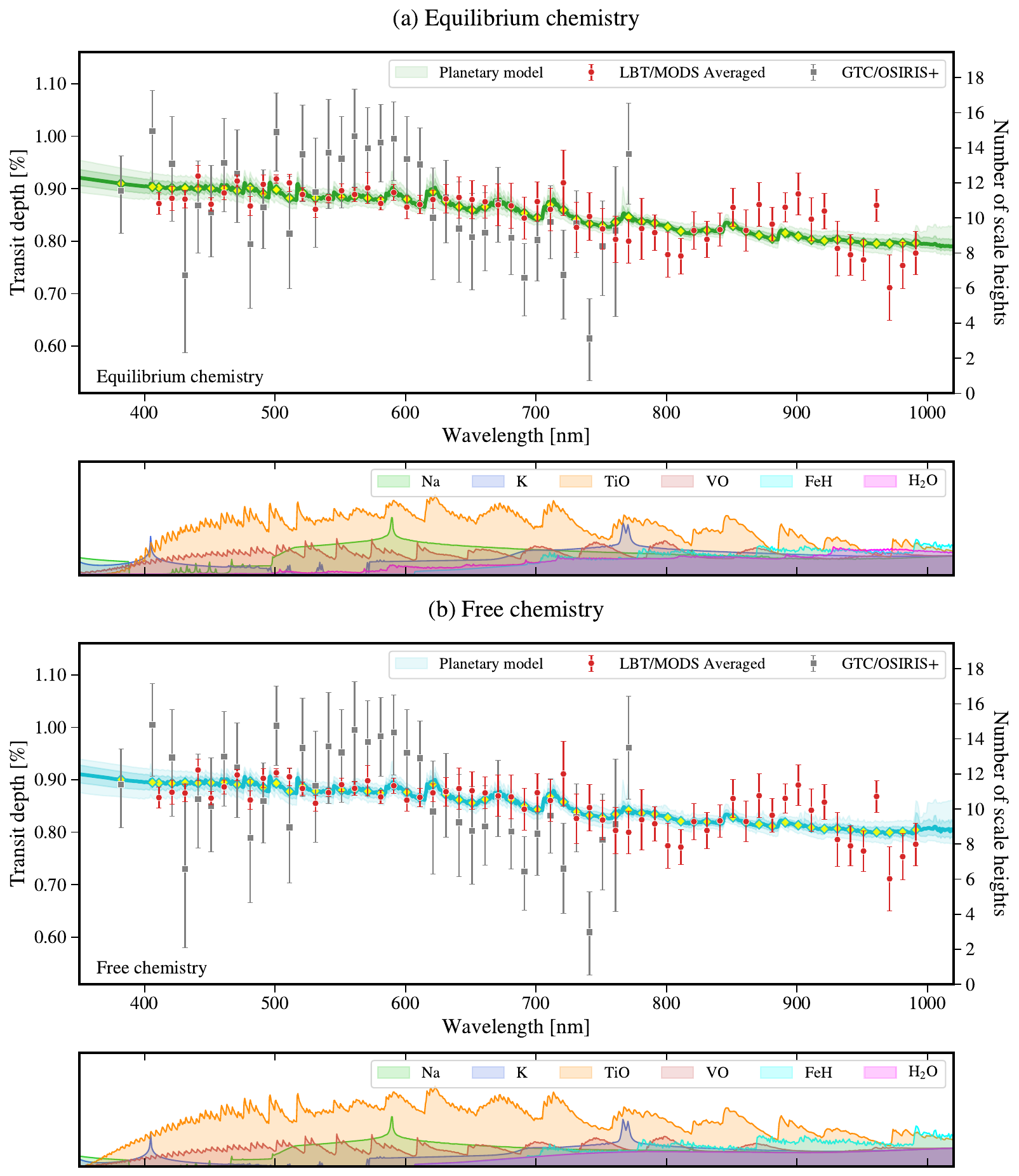}
\caption{Joint retrieval of the LBT/MODS and GTC/OSIRIS+ transmission spectra of HAT-P-47b assuming equilibrium chemistry (panel a) and free chemistry (panel b). Top sub-panels: observed transmission spectra with the best-fit models, after applying the retrieved vertical offsets and error-scaling factors. The shaded regions indicate the median model and the associated $1\sigma$ and $2\sigma$ confidence intervals. Bottom sub-panels: reference models illustrating the individual contributions of key opacity species. }
\label{fig:retrieval_Joint}
\end{figure*}

\begin{figure}%[htb]
\centering
\includegraphics[width=\hsize]{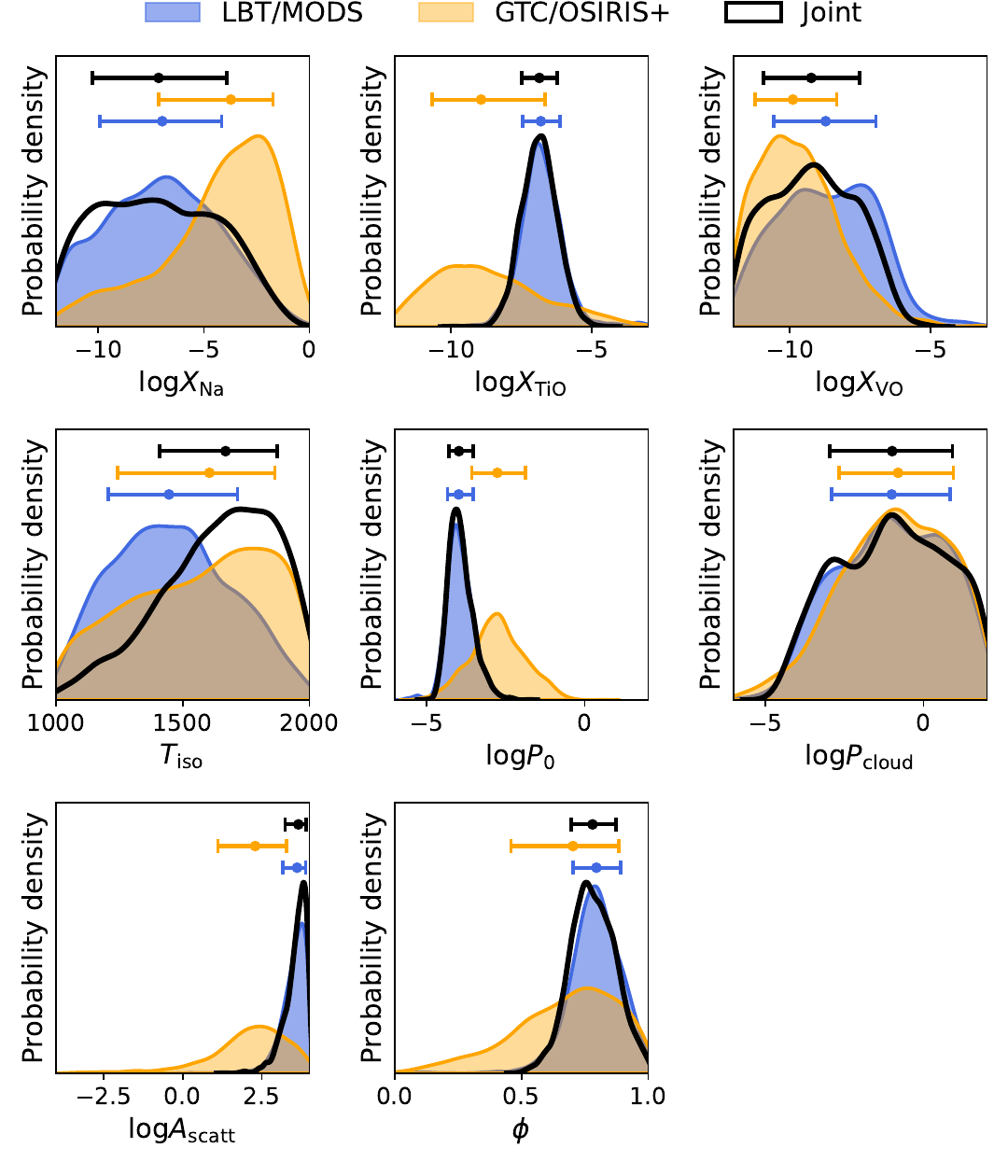}
\caption{Posterior distributions of the retrieved atmospheric parameters assuming free chemistry. Results from the LBT/MODS retrieval (blue), GTC/OSIRIS+ retrieval (orange), and the joint retrieval (black) are shown for comparison. Horizontal error bars indicate the median values and corresponding 1$\sigma$ credible intervals. Species with unconstrained abundances are not shown. }
\label{fig:post_cmp}
\end{figure}

\subsubsection{Retrieval results}
The hybrid model incorporating stellar contamination is disfavored relative to the pure atmospheric scenario ($\Delta\ln\mathcal{Z}=-4.31$) and requires unrealistically large stellar heterogeneity. The inferred spot and facular covering fractions ($38^{+12}_{-8}$\% and $20^{+19}_{-12}$\% for MODS; $50^{+13}_{-12}$\% and $13^{+13}_{-9}$\% for OSIRIS+) are inconsistent with expectations for such a quiet star and correspond to a $\sim$$5.7^{+3.1}_{-4.5}$\% stellar variability between the two observing epochs. We further examined the TESS photometry and found that the transit depths are consistent across multiple sectors within uncertainties: $R_\mathrm{p}/R_\star = 0.08909^{+0.00090}_{-0.00084}$ in S18 (2019), $0.08901^{+0.00083}_{-0.00080}$ in S58 (2022), and $0.08825^{+0.00080}_{-0.00081}$ in S85 (2024). Although a stellar variability of $\sim$$1.9^{+2.8}_{-2.1}$\% between S18 and S85 cannot be ruled out by the TESS transit depths alone, such variability levels are an order of magnitude larger than the amplitudes measured by Kepler for stars with $T_\mathrm{eff}=6425$--6575~K \citep[$0.12^{+0.19}_{-0.05}$\%;][]{McQuillan2014,Rackham2019} and by TESS for stars with 6650--6750~K \citep[$0.12^{+0.36}_{-0.06}$\%;][]{Boyle2026}. We therefore consider stellar contamination unlikely to be the dominant origin of the observed transmission spectral features and focus the following analysis primarily on pure atmospheric interpretations.

We present the retrieval results for the separate LBT/MODS and GTC/OSIRIS+ datasets and for the joint analysis. The retrieved parameters are listed in Table~\ref{tab:retrieved_param}, and model comparison statistics are summarized in Table~\ref{tab:model_cmp}. The best-fit spectra for the joint datasets and posterior distributions of key parameters are shown in Figs.~\ref{fig:retrieval_Joint} and \ref{fig:post_cmp}, while the results for the individual datasets are provided in the Appendix (Figs.~\ref{fig:retrieval_MODS} and \ref{fig:retrieval_OSIRIS}). 

According to Bayesian model comparison (Table~\ref{tab:model_cmp}), atmospheric models are weakly favored over a simple sloped model for MODS, statistically indistinguishable for OSIRIS+, and moderately favored for the joint dataset. In contrast, they are moderately favored for OSIRIS+ and strongly favored for both MODS and the joint dataset when compared to a simple flat model.  

The equilibrium and free chemistry retrievals yield broadly consistent parameter estimates for the MODS-only and joint datasets. In particular, the atmospheric temperature, reference pressure, cloud properties, dataset-dependent offsets, and error scaling factors agree within the $1\sigma$ level. In contrast, the OSIRIS+-only retrieval shows larger discrepancies between the two chemical parameterizations, with median values differing by up to $1.6\sigma$. The equilibrium-chemistry retrieval for OSIRIS+ provides only weak constraints on several parameters. The posterior distribution of the atmospheric temperature $T_{\rm{iso}}$ is bimodal and exhibits degeneracy with the reference pressure $\log P_0$, which is further correlated with other atmospheric parameters, including the scattering amplitude $\log A_{\mathrm{RS}}$, cloud fraction $\phi$, and metallicity $\log Z$. 

The retrieved atmospheric temperature is broadly consistent across all analyses. The joint retrieval yields $T_{\rm iso}\sim 1400-1900$~K, consistent with the equilibrium temperature ($T_{\rm eq}=1605$~K) within uncertainties. The MODS-only and OSIRIS+-only retrievals show broader posteriors, with a tendency toward lower temperatures. The reference pressure is moderately constrained in the joint retrieval, with $\log P_0 = -3.42^{+0.51}_{-0.37}$~$\log {\rm bar}$ (equilibrium chemistry) and $-3.98^{+0.45}_{-0.31}$~$\log {\rm bar}$ (free chemistry), corresponding to pressures of $\sim 10^{-4}-10^{-3}$~bar. Constraints from the OSIRIS+-only retrieval are weaker. 

Cloud properties are consistent across datasets and chemical assumptions. The cloud coverage fraction is $\phi \sim$0.7--0.9, indicating that a large portion of the terminator is affected by clouds. The cloud-top pressure is only loosely constrained but generally lies in the millibar-to-decibar range. For the MODS and joint retrievals, the scattering enhancement factor is skewed toward the upper prior boundary, with $\log A_\mathrm{RS} > 3.09$ ($90\%$ lower limit, joint free chemistry case), suggesting additional short-wavelength aerosol opacity beyond molecular hydrogen Rayleigh scattering. In contrast, the parameter remains weakly constrained in the OSIRIS+-only retrieval. 

Despite the large inferred cloud coverage fraction, the transmission spectra still exhibit wavelength-dependent structures beyond a purely featureless sloped spectrum. The retrieved cloud-top pressures in the millibar-to-decibar range suggest that the aerosol layer does not completely obscure the observable atmosphere, allowing the transmission spectra to probe lower-opacity regions above at least part of the cloud deck. This is likely aided by the large atmospheric scale height of HAT-P-47b. A comparable example is GJ~1214b, whose transmission spectrum was long interpreted as being strongly muted by high-altitude aerosols until recent JWST observations revealed weak CO$_2$ and CH$_4$ features above the aerosol continuum \citep{Schlawin2024}. In the case of HAT-P-47b, the combination of low surface gravity and extended atmosphere may similarly facilitate transmission-spectroscopy measurements above cloudy regions despite substantial aerosol coverage. High-resolution spectroscopy with facilities such as ESPRESSO or MAROON-X could probe the upper atmospheric layers above the cloud deck \citep[e.g.,][]{Gandhi2023,Prinoth2023}.

Under the equilibrium chemistry assumption, the joint retrieval yields a C/O ratio of $0.62^{+0.37}_{-0.35}$ and a metallicity constrained toward the lower prior boundary ($\log Z < 0.53$, $90\%$ upper limit), with both parameters remaining weakly constrained. Similar trends are found for the MODS-only retrieval, while the OSIRIS+-only data do not provide meaningful constraints. 

In the free chemistry retrieval, only a subset of species is constrained, with most abundances remaining prior-dominated. The wiggle-like absorption and overall slope are primarily reproduced by TiO opacity combined with aerosol scattering. The posterior distributions (Fig.~\ref{fig:post_cmp}) show that TiO is well constrained in the MODS and joint retrievals, with $\log X_\mathrm{TiO} = -6.81^{+0.69}_{-0.65}$ and $-6.86^{+0.64}_{-0.63}$, respectively. The OSIRIS+-only retrieval yields a weaker constraint ($\log X_\mathrm{TiO} = -8.94^{+2.27}_{-1.73}$), but remains consistent within uncertainties. VO is weakly constrained in all cases. A marginal constraint on Na is obtained only for the OSIRIS+ dataset. 

To assess the contribution of individual opacity sources, we performed additional retrievals in which each of the 15 species was removed in turn from the free chemistry model. The resulting Bayesian evidence (Table~\ref{tab:model_cmp}) shows that TiO is the only species whose removal leads to a moderate decrease. Excluding TiO reduces the evidence by $\Delta \ln \mathcal{Z} = 2.68$ for the MODS-only retrieval and by 3.44 for the joint retrieval, corresponding to moderate evidence against TiO-free models. Removing other species produces negligible changes in the evidence.

\section{Discussion}
\label{Section:discussion}
\subsection{Enhanced scattering from aerosols}
Both the MODS and OSIRIS+ datasets exhibit a general decrease in transit depth toward longer wavelengths, with only weak additional spectral signatures. Our joint retrieval indicates that this behavior can be explained by enhanced aerosol scattering, corresponding to a scattering amplitude more than three orders of magnitude stronger than H$_2$ Rayleigh scattering. 

Sloped optical transmission spectra are commonly observed in hot Jupiters. In some cases, such slopes can be attributed to stellar contamination from unocculted spots or faculae \citep[e.g., HAT-P-18b;][]{Fournier-Tondreau2024}. In other systems, synergistic optical and infrared observations have demonstrated that the slopes originate from high-altitude aerosols composed of submicron particles, whose presence is supported by infrared vibrational features of aerosol species detected in transmission \citep[e.g., WASP-17b;][]{Alderson2022,Grant2023,Louie2025} or emission spectra \citep[e.g., HD 189733b;][]{Pont2013,Mullens2024,Inglis2024}. 

In the case of HAT-P-47b, the stellar contamination retrievals are statistically disfavored and require unrealistically large levels of stellar heterogeneity. Therefore, the observed optical slope is most likely produced by aerosol scattering in its atmosphere. Although these aerosols generally lack distinctive features in the optical, infrared observations will be important for constraining their compositions and particle properties. In particular, observations with the JWST could probe vibrational signatures of cloud species and help determine the nature of the aerosols in HAT-P-47b's atmosphere \citep[e.g.][]{Grant2023,Dyrek2024,Bell2024}.

\subsection{Tentative evidence of TiO}
Our joint retrieval of the MODS and OSIRIS+ datasets reveals that the weak spectral features superimposed on the overall slope can be tentatively explained by the TiO absorption. 

Although our inference of TiO in the atmosphere of HAT-P-47b is only tentative, if present, TiO would be particularly interesting given the planet's equilibrium temperature ($T_{\rm eq}=1605$~K), which lies close to the expected condensation boundary of Ti-bearing species \citep{Lodders2003,Sharp2007,Fortney2008}. The TiO mass fractions inferred from our free-chemistry retrievals appear consistent with those derived from the equilibrium-chemistry retrieval ($\log X_\mathrm{TiO} = -7.15^{+1.10}_{-0.69}$ for $P<0.1$~bar), suggesting that the retrieved abundances are chemically plausible within the adopted equilibrium framework. However, we note that the equilibrium-chemistry prescription adopted in \texttt{petitRADTRANS} is based on chemical tables generated with \texttt{easyCHEM} \citep{2017A&A...600A..10M,2025JOSS...10.7712L}, which include equilibrium condensation but do not account for vertical elemental depletion caused by condensate rainout. When rainout is considered from high- to low-pressure layers, Ti-bearing species can condense in deeper atmospheric layers, thereby reducing the elemental Ti inventory available at lower pressures and strongly suppressing gas-phase TiO abundances at observable altitudes, unless the deep atmosphere remains sufficiently hot to prevent condensation. In this context, the possible presence of TiO in HAT-P-47b may imply a non-isothermal temperature structure in which the deeper atmosphere stays above the condensation temperature of Ti-bearing species over the relevant pressure range.

On the other hand, gas-phase TiO may remain observable over a broader temperature range if vertical mixing replenishes Ti-bearing gas above the condensation level, or if vigorous day–night circulation sustains the observed abundance from competition between atmospheric transport and nightside cold trapping \citep{Showman2009,Parmentier2013,Parmentier2018}. Such mechanisms may be particularly important near the expected condensation boundary of Ti-bearing species, where small differences in atmospheric dynamics or temperature structure could lead to substantial differences in the observable TiO abundance.

The relatively muted absorption signatures in the current spectrum may also reflect longitudinal variations at the planetary terminator. Three-dimensional circulation models predict that the morning and evening limbs can exhibit different cloud properties and chemical abundances. A comparable case is WASP-94b \citep{Ahrer2025}, which has a similar temperature and surface gravity (1604~K and 3.48~$\rm{m~s^{-2}}$, versus 1605~K and 2.95~$\rm{m~s^{-2}}$ for HAT-P-47b) and shows evidence for cloudy morning and clear evening limbs shaped by atmospheric circulation. A similar mechanism could potentially reduce the net absorption amplitude in the disk-integrated spectrum of HAT-P-47b. Future observations with the JWST will be important for testing this scenario. 

Higher-precision spectroscopy over a broader wavelength range could help confirm the presence of TiO and place stronger constraints on the role of atmospheric circulation in shaping the transmission spectrum of HAT-P-47b. In particular, high-resolution spectroscopy may probe atmospheric layers above the cloud deck that are less accessible to low-resolution transmission spectroscopy, providing a complementary view of the upper atmosphere and potentially enhancing sensitivity to Ti-bearing species despite substantial cloud coverage \citep{Pino2018}. Some early high-resolution spectroscopic claims of TiO were affected by imperfect molecular line lists \citep[e.g.,][]{Hoeijmakers2015}. Subsequent improvements in TiO opacity databases have substantially improved the reliability of high-resolution analyses and enabled robust detections in systems such as WASP-189b \citep{Prinoth2022}.

If TiO is indeed present in the upper atmosphere, it may efficiently absorb incoming stellar radiation and potentially lead to a thermal inversion in the temperature–pressure profile. In this context, secondary-eclipse emission spectroscopy would provide an important test of the role of TiO opacity in shaping the atmospheric structure.

\section{Conclusions}
\label{Section:conclusions}
HAT-P-47b is a low-mass, highly inflated hot sub-Saturn located near the boundary of the short-period Neptunian desert. With an equilibrium temperature of $T_{\rm eq}=1605\pm22$~K \citep{Bakos2016}, it lies close to the predicted condensation threshold of TiO species, making it a compelling target for investigating the presence of optical absorbers and their potential role in shaping atmospheric temperature structures. 

Using 13 transits observed by TESS, we refined the orbital ephemeris and transit parameters of HAT-P-47b, significantly improving the precision of the orbital period and reference epoch. We then presented two independent ground-based optical transmission spectra obtained with LBT/MODS and GTC/OSIRIS+, corresponding to two separate transits. 

The analyses of the two spectra reveal distinct spectral behaviors. The higher-precision MODS spectrum shows moderate evidence for the presence of TiO absorption, while the OSIRIS+ spectrum does not provide statistically significant support. After accounting for inter-channel normalization offsets, both datasets require at least a moderate scattering enhancement factor to reproduce the observed optical slope, indicating the presence of additional aerosol opacity beyond pure molecular Rayleigh scattering. The joint retrieval is primarily driven by the higher signal-to-noise MODS data and yields moderate evidence for TiO absorption, while also providing a meaningful constraint on its abundance. 

Given the limited spectral resolution and measurement precision of ground-based low-resolution optical transmission spectra, conclusively confirming the presence of TiO remains challenging. Future high-resolution spectroscopy and high-precision space-based observations with HST and JWST \citep[e.g.,][]{Edwards2024} would provide improved constraints on both aerosol properties and chemical abundances, potentially offering insights into the planet's formation and migration pathways.

\begin{acknowledgements}
We thank the anonymous referees for their constructive comments and suggestions. G.C. acknowledges the support by the National Natural Science Foundation of China (NSFC grant Nos. 42578016, 12122308, 42075122), Youth Innovation Promotion Association CAS (2021315), and the Minor Planet Foundation of the Purple Mountain Observatory. 
This work is partly based on observations made with the Gran Telescopio Canarias installed at the Spanish Observatorio del Roque de los Muchachos of the Instituto de Astrofísica de Canarias on the island of La Palma. 
This work is partly based on observations made with the Large Binocular Telescope. The LBT is an international collaboration among institutions in the United States, Italy and Germany. LBT Corporation partners are: The University of Arizona on behalf of the Arizona Board of Regents; Istituto Nazionale di Astrofisica, Italy; LBT Beteiligungsgesellschaft, Germany, representing the Max-Planck Society, The Leibniz Institute for Astrophysics Potsdam, and Heidelberg University; The Ohio State University, representing OSU, University of Notre Dame, University of Minnesota and University of Virginia. We acknowledge the support from the LBT-Italian Coordination Facility for the execution of observations, data distribution and reduction.
This paper made use of the modsCCDRed data reduction code developed in part with funds provided by NSF Grants AST-9987045 and AST-1108693. 
\end{acknowledgements}

\bibliographystyle{aa.bst} 
\bibliography{reference}

\begin{appendix}

\onecolumn
\section{Additional tables and figures}

\begin{longtable}{ccccc}
\caption{Derived transmission spectra of HAT-P-47b for each spectroscopic channel.}\\
\renewcommand\arraystretch{1.6}
\label{tab:spectra} 
\endfirsthead
\caption{continued.}\\
\hline
\multirow{2}{*}{Wavelength~[nm]} & \multicolumn{4}{c}{$R_{\rm{p}}/R_\star$} \\\noalign{\smallskip}
& LBT/MODS1B & LBT/MODS1R & LBT/MODS2B & GTC/OSIRIS+\\\noalign{\smallskip}
\hline
\endhead
\hline
\endfoot
\hline
\multirow{2}{*}{Wavelength~[nm]} & \multicolumn{4}{c}{$R_{\rm{p}}/R_\star$} \\\noalign{\smallskip}
& LBT/MODS1B & LBT/MODS1R & LBT/MODS2B & GTC/OSIRIS+\\\noalign{\smallskip}
\hline
368--396 & -- & -- & -- & $0.0947^{+0.0060}_{-0.0073}$ \\\noalign{\smallskip}
396--416 & -- & -- & -- & $0.1006^{+0.0066}_{-0.0082}$ \\\noalign{\smallskip}
406--416 & $0.0884^{+0.0019}_{-0.0018}$ & -- & $0.0890^{+0.0025}_{-0.0024}$ & -- \\\noalign{\smallskip}
416--426 & $0.0898^{+0.0051}_{-0.0030}$ & -- & $0.0879^{+0.0021}_{-0.0016}$ & $0.0974^{+0.0079}_{-0.0096}$ \\\noalign{\smallskip}
426--436 & $0.0883^{+0.0025}_{-0.0016}$ & -- & $0.0886^{+0.0016}_{-0.0014}$ & $0.0858^{+0.0159}_{-0.0147}$ \\\noalign{\smallskip}
436--446 & $0.0919^{+0.0021}_{-0.0021}$ & -- & $0.0906^{+0.0019}_{-0.0020}$ & $0.0932^{+0.0078}_{-0.0083}$ \\\noalign{\smallskip}
446--456 & $0.0910^{+0.0022}_{-0.0018}$ & -- & $0.0867^{+0.0011}_{-0.0010}$ & $0.0926^{+0.0083}_{-0.0079}$ \\\noalign{\smallskip}
456--466 & $0.0911^{+0.0013}_{-0.0014}$ & -- & $0.0879^{+0.0014}_{-0.0012}$ & $0.0975^{+0.0074}_{-0.0081}$ \\\noalign{\smallskip}
466--476 & $0.0913^{+0.0013}_{-0.0013}$ & -- & $0.0900^{+0.0010}_{-0.0010}$ & $0.0964^{+0.0074}_{-0.0083}$ \\\noalign{\smallskip}
476--486 & $0.0892^{+0.0018}_{-0.0016}$ & -- & $0.0870^{+0.0021}_{-0.0018}$ & $0.0892^{+0.0105}_{-0.0116}$ \\\noalign{\smallskip}
486--496 & $0.0919^{+0.0022}_{-0.0018}$ & -- & $0.0890^{+0.0016}_{-0.0018}$ & $0.0931^{+0.0066}_{-0.0072}$ \\\noalign{\smallskip}
496--506 & $0.0920^{+0.0012}_{-0.0012}$ & -- & $0.0899^{+0.0007}_{-0.0006}$ & $0.1005^{+0.0063}_{-0.0063}$ \\\noalign{\smallskip}
506--516 & $0.0916^{+0.0019}_{-0.0020}$ & -- & $0.0895^{+0.0016}_{-0.0014}$ & $0.0903^{+0.0104}_{-0.0099}$ \\\noalign{\smallskip}
516--526 & $0.0904^{+0.0011}_{-0.0011}$ & -- & $0.0883^{+0.0017}_{-0.0014}$ & $0.0983^{+0.0081}_{-0.0079}$ \\\noalign{\smallskip}
526--536 & $0.0890^{+0.0014}_{-0.0013}$ & -- & $0.0863^{+0.0020}_{-0.0019}$ & $0.0946^{+0.0093}_{-0.0095}$ \\\noalign{\smallskip}
536--546 & $0.0905^{+0.0018}_{-0.0019}$ & -- & $0.0877^{+0.0006}_{-0.0006}$ & $0.0985^{+0.0088}_{-0.0093}$ \\\noalign{\smallskip}
546--556 & $0.0910^{+0.0012}_{-0.0011}$ & -- & $0.0883^{+0.0017}_{-0.0014}$ & $0.0979^{+0.0070}_{-0.0082}$ \\\noalign{\smallskip}
556--566 & $0.0900^{+0.0020}_{-0.0020}$ & -- & $0.0884^{+0.0012}_{-0.0009}$ & $0.1001^{+0.0077}_{-0.0083}$ \\\noalign{\smallskip}
566--576 & -- & $0.0948^{+0.0035}_{-0.0033}$ & $0.0890^{+0.0030}_{-0.0023}$ & $0.0989^{+0.0067}_{-0.0072}$ \\\noalign{\smallskip}
576--586 & -- & $0.0925^{+0.0025}_{-0.0024}$ & $0.0874^{+0.0011}_{-0.0009}$ & $0.0995^{+0.0062}_{-0.0065}$ \\\noalign{\smallskip}
586--596 & -- & $0.0950^{+0.0022}_{-0.0022}$ & $0.0883^{+0.0014}_{-0.0010}$ & $0.0998^{+0.0061}_{-0.0068}$ \\\noalign{\smallskip}
596--606 & -- & $0.0940^{+0.0030}_{-0.0026}$ & $0.0864^{+0.0016}_{-0.0022}$ & $0.0979^{+0.0070}_{-0.0075}$ \\\noalign{\smallskip}
606--616 & -- & $0.0943^{+0.0020}_{-0.0022}$ & $0.0865^{+0.0015}_{-0.0018}$ & $0.0974^{+0.0061}_{-0.0067}$ \\\noalign{\smallskip}
616--626 & -- & $0.0935^{+0.0029}_{-0.0026}$ & $0.0878^{+0.0015}_{-0.0013}$ & $0.0920^{+0.0088}_{-0.0109}$ \\\noalign{\smallskip}
626--636 & -- & $0.0929^{+0.0031}_{-0.0033}$ & $0.0884^{+0.0026}_{-0.0023}$ & $0.0941^{+0.0063}_{-0.0079}$ \\\noalign{\smallskip}
636--646 & -- & $0.0940^{+0.0028}_{-0.0028}$ & -- & $0.0909^{+0.0084}_{-0.0096}$ \\\noalign{\smallskip}
646--656 & -- & $0.0938^{+0.0026}_{-0.0026}$ & -- & $0.0899^{+0.0080}_{-0.0095}$ \\\noalign{\smallskip}
656--666 & -- & $0.0935^{+0.0024}_{-0.0020}$ & -- & $0.0904^{+0.0071}_{-0.0069}$ \\\noalign{\smallskip}
666--676 & -- & $0.0932^{+0.0031}_{-0.0027}$ & -- & $0.0935^{+0.0062}_{-0.0068}$ \\\noalign{\smallskip}
676--686 & -- & $0.0931^{+0.0033}_{-0.0028}$ & -- & $0.0899^{+0.0059}_{-0.0067}$ \\\noalign{\smallskip}
686--696 & -- & $0.0919^{+0.0031}_{-0.0027}$ & -- & $0.0855^{+0.0066}_{-0.0072}$ \\\noalign{\smallskip}
696--706 & -- & $0.0936^{+0.0026}_{-0.0027}$ & -- & $0.0896^{+0.0069}_{-0.0074}$ \\\noalign{\smallskip}
706--716 & -- & $0.0928^{+0.0031}_{-0.0028}$ & -- & $0.0916^{+0.0065}_{-0.0065}$ \\\noalign{\smallskip}
716--726 & -- & $0.0955^{+0.0043}_{-0.0045}$ & -- & $0.0859^{+0.0085}_{-0.0084}$ \\\noalign{\smallskip}
726--736 & -- & $0.0909^{+0.0035}_{-0.0035}$ & -- & $0.0914^{+0.0057}_{-0.0062}$ \\\noalign{\smallskip}
736--746 & -- & $0.0920^{+0.0035}_{-0.0031}$ & -- & $0.0785^{+0.0083}_{-0.0087}$ \\\noalign{\smallskip}
746--756 & -- & $0.0907^{+0.0031}_{-0.0026}$ & -- & $0.0890^{+0.0082}_{-0.0090}$ \\\noalign{\smallskip}
756--766 & -- & $0.0896^{+0.0033}_{-0.0033}$ & -- & $0.0907^{+0.0115}_{-0.0155}$ \\\noalign{\smallskip}
766--776 & -- & $0.0894^{+0.0028}_{-0.0034}$ & -- & $0.0984^{+0.0084}_{-0.0084}$ \\\noalign{\smallskip}
776--786 & -- & $0.0908^{+0.0050}_{-0.0037}$ & -- & -- \\\noalign{\smallskip}
786--796 & -- & $0.0903^{+0.0025}_{-0.0025}$ & -- & -- \\\noalign{\smallskip}
796--806 & -- & $0.0880^{+0.0032}_{-0.0034}$ & -- & -- \\\noalign{\smallskip}
806--816 & -- & $0.0878^{+0.0026}_{-0.0026}$ & -- & -- \\\noalign{\smallskip}
816--826 & -- & $0.0906^{+0.0026}_{-0.0026}$ & -- & -- \\\noalign{\smallskip}
826--836 & -- & $0.0896^{+0.0024}_{-0.0028}$ & -- & -- \\\noalign{\smallskip}
836--846 & -- & $0.0907^{+0.0023}_{-0.0024}$ & -- & -- \\\noalign{\smallskip}
846--856 & -- & $0.0930^{+0.0023}_{-0.0030}$ & -- & -- \\\noalign{\smallskip}
856--866 & -- & $0.0906^{+0.0030}_{-0.0028}$ & -- & -- \\\noalign{\smallskip}
866--876 & -- & $0.0933^{+0.0027}_{-0.0034}$ & -- & -- \\\noalign{\smallskip}
876--886 & -- & $0.0912^{+0.0025}_{-0.0022}$ & -- & -- \\\noalign{\smallskip}
886--896 & -- & $0.0930^{+0.0016}_{-0.0025}$ & -- & -- \\\noalign{\smallskip}
896--906 & -- & $0.0943^{+0.0022}_{-0.0034}$ & -- & -- \\\noalign{\smallskip}
906--916 & -- & $0.0918^{+0.0025}_{-0.0035}$ & -- & -- \\\noalign{\smallskip}
916--926 & -- & $0.0926^{+0.0019}_{-0.0030}$ & -- & -- \\\noalign{\smallskip}
926--936 & -- & $0.0887^{+0.0035}_{-0.0043}$ & -- & -- \\\noalign{\smallskip}
936--946 & -- & $0.0880^{+0.0030}_{-0.0029}$ & -- & -- \\\noalign{\smallskip}
946--956 & -- & $0.0874^{+0.0027}_{-0.0033}$ & -- & -- \\\noalign{\smallskip}
956--966 & -- & $0.0932^{+0.0022}_{-0.0022}$ & -- & -- \\\noalign{\smallskip}
966--976 & -- & $0.0843^{+0.0047}_{-0.0052}$ & -- & -- \\\noalign{\smallskip}
976--986 & -- & $0.0868^{+0.0027}_{-0.0041}$ & -- & -- \\\noalign{\smallskip}
986--996 & -- & $0.0882^{+0.0024}_{-0.0038}$ & -- & -- \\\noalign{\smallskip}
\hline
\end{longtable}

\begin{figure*}[hb]
\centering
\includegraphics[width=0.85\hsize]{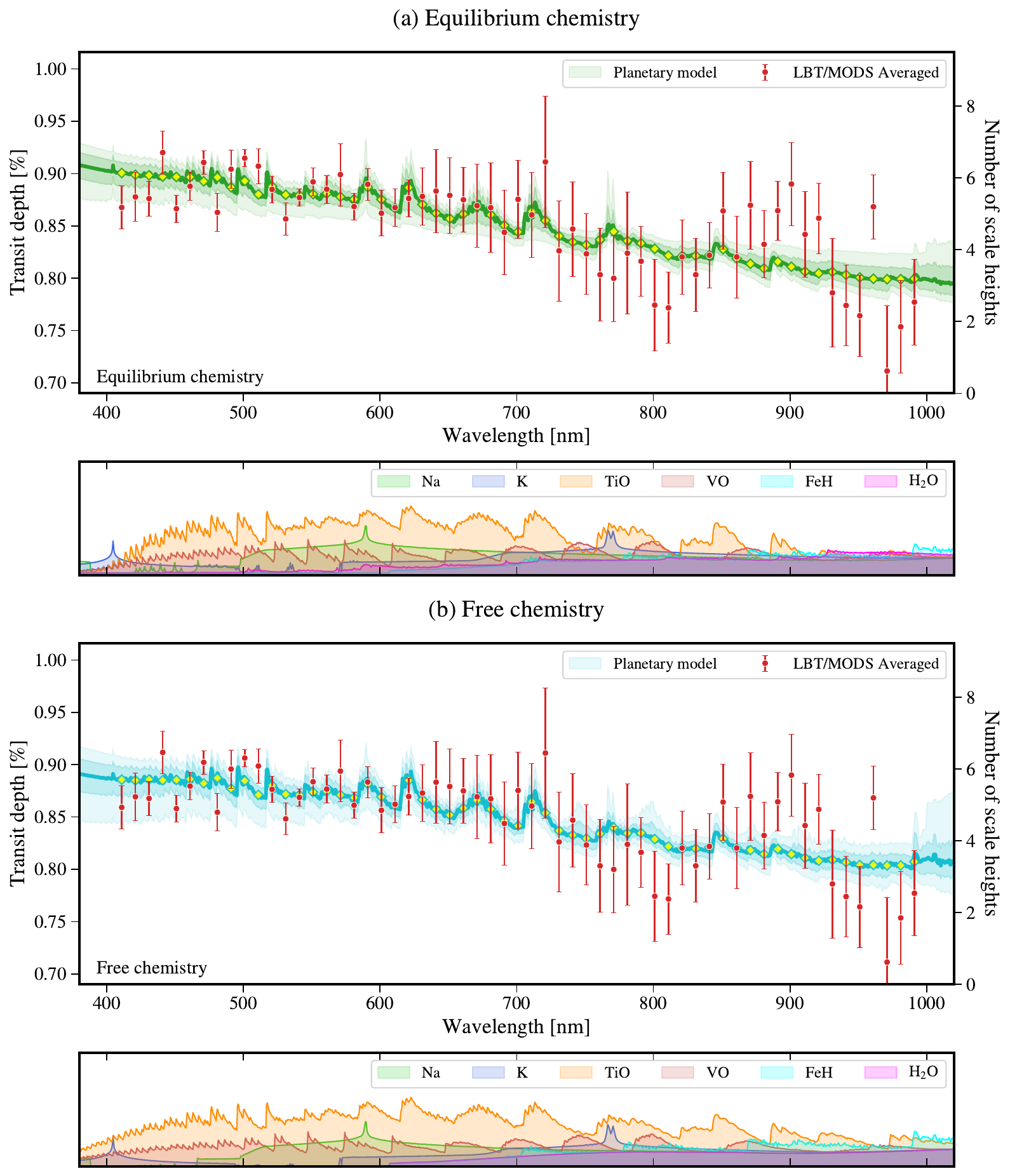}
\caption{Same as Fig.~\ref{fig:retrieval_Joint}, but for the LBT/MODS transmission spectra. }
\label{fig:retrieval_MODS}
\end{figure*}

\begin{figure*}[hb]
\centering
\includegraphics[width=0.85\hsize]{{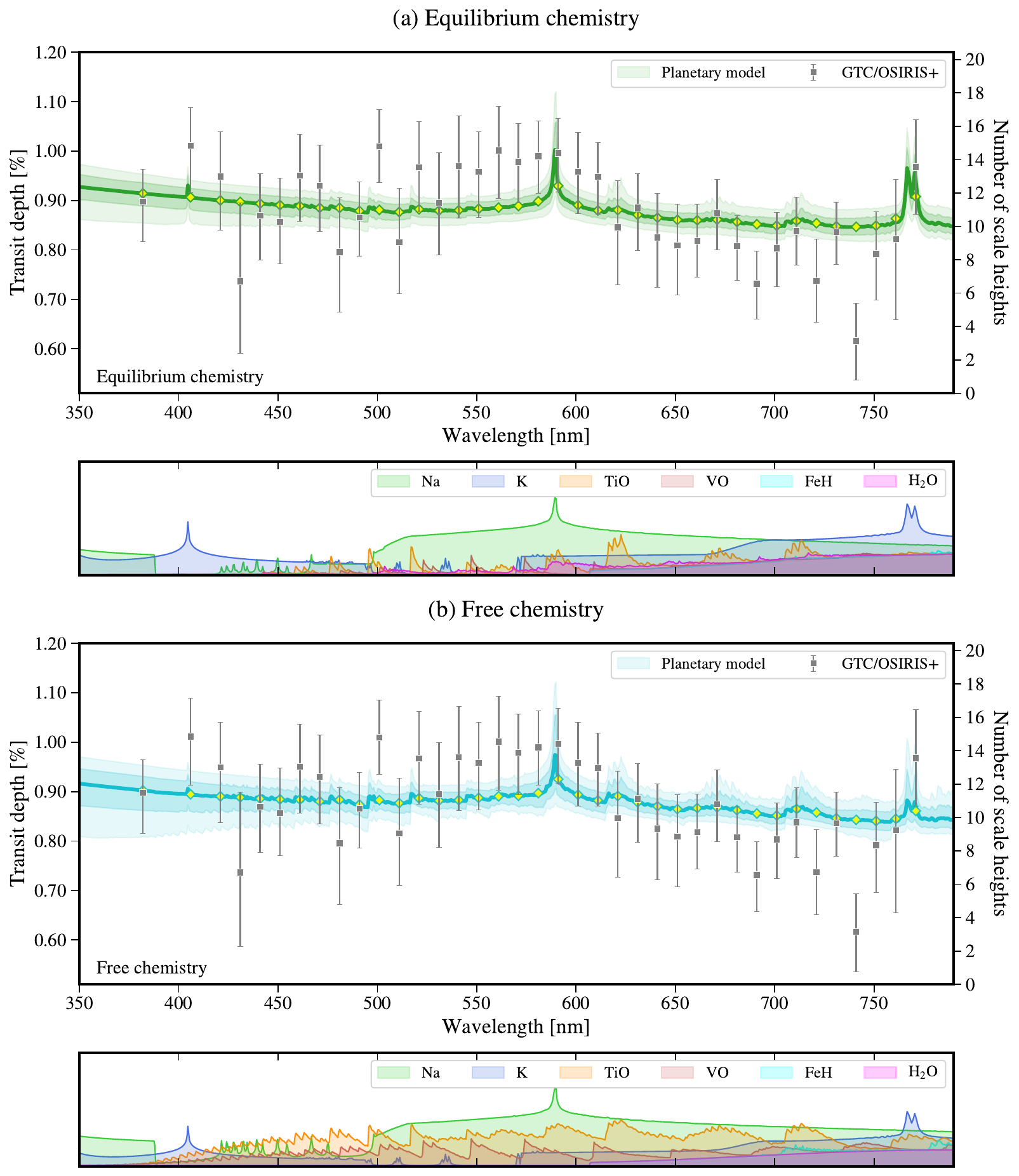}}
\caption{Same as Fig.~\ref{fig:retrieval_MODS}, but for the GTC/OSIRIS+ transmission spectrum. }
\label{fig:retrieval_OSIRIS}
\end{figure*}

\FloatBarrier 
\twocolumn
\FloatBarrier 
\clearpage

\end{appendix}
\end{document}